\documentclass{aa}  
\usepackage{natbib}
\bibpunct{(}{)}{;}{a}{}{,} 
\usepackage{graphicx}
\usepackage{adjustbox} 
\usepackage{subcaption}
\usepackage{verbatim}
\usepackage[varg]{txfonts}
\usepackage{setspace}
\usepackage{booktabs}
\usepackage{multirow}
\usepackage{comment}
\usepackage{float}
\usepackage{tikz}
\usepackage{color}
\graphicspath{{./}{./figure/}}
\usepackage{xcolor}
\usepackage{placeins}
\usepackage{stfloats}
\usepackage[colorlinks=true,linkcolor=blue, citecolor=blue, filecolor=blue, urlcolor=blue]{hyperref}

\usetikzlibrary{arrows,positioning} 
\tikzset{
    >=stealth',
    punkt/.style={
           rectangle,
           rounded corners,
           draw=black, very thick,
           text width=6.5em,
           minimum height=2em,
           text centered},
    pil/.style={
           ->,
           thick,
           shorten <=2pt,
           shorten >=2pt,}
}
\newcommand\rectagular[1][red]{\begin{tikzpicture}
\draw [fill=red,red] (0.2,0.2) rectangle (0.3,0.3); 
\end{tikzpicture}
}

\begin{document} 

    \title{A Remote-Sensing Framework for Interpreting CME Transit Effects}
    \author{S.Stamkos\inst{1} 
    \and A. Vourlidas\inst{2}
    \and E. Paouris\inst{2}
    \and S.Patsourakos\inst{1}
  }
\institute{Department of Physics, Section of Astrogeophysics, University of Ioannina, 45110, Greece. 
\and Applied Physics Laboratory, Johns Hopkins University, Laurel, MD 20723, USA
}
    
\date{Received date / Accepted date }

\abstract {}{This proof-of-concept study develops a remote-sensing methodology to identify CME transit effects in the inner heliosphere, using an ideal self-similar reference shell together with a compression model to place observed density falloffs in context. We apply the method to assess how pile-up shapes the density structure, and to what extent weaker, localized departures, potentially consistent with magnetic erosion, can be isolated once compression and line-of-sight projection effects are accounted for.}{We analyze the 27 January 2012 CME using STEREO/SECCHI COR2-A, together with COR2-B and SOHO/LASCO C3. Observed electron-density profiles are extracted along the CME front in position-angle sectors and compared with synthetic profiles from a forward-modeled, self-similar GCS shell. The synthetic profiles are normalized to the observed density at a selected reference height, establishing a common baseline for comparison. We quantify radial density evolution using sector-wise power-law fits that account for measurement scatter. These trends are compared with a pile-up baseline, and residual differences are used to identify localized departures.}{Sector-wise density indices reveal nonuniform evolution across the CME front. In COR2-A, western sectors exhibit pronounced flattening relative to the synthetic shell, approaching the snowplow-like compression benchmark, while several lower and central PA sectors show steeper-than-synthetic falloffs. COR2-B and LASCO C3 also show sector-dependent departures, with PA patterns varying with viewing geometry. Using a conservative excess criterion, clear pile-up is detected only in COR2-A sectors at each time step, indicating spatially structured compression. Snowplow comparisons depend strongly on the assumed upstream wind: the fast-stream case shows no density deficit, whereas the slow-stream case yields deficits across all PA sectors. These results indicate that compression alone is insufficient to explain the observed evolution, which combines compression-dominated regions with localized steepening potentially reflecting depletion, material redistribution, or erosion-like evolution.}{This study shows that coronagraph-derived CME-front density profiles can be used as a quantitative diagnostic by comparing an idealized self-similar expansion reference with snowplow-based compression benchmarks. Applied to the 27 January 2012 CME, the results show a mixed front rather than a single global behavior: some sectors are compression-dominated and flatten toward the snowplow-like benchmark, while others fall off more steeply than the synthetic shell and therefore identify candidate regions of depletion or erosion-like evolution. Applied to larger samples and combined with event-local in situ constraints and modeling, this framework provides a practical route to separate the physical processes shaping CME evolution during transit.}

\keywords{Coronal Mass Ejections (CMEs) -- Space Weather -- Pile-up -- Magnetic Erosion -- Remote-Sensing Observations -- Synthetic Images -- Data Analysis} 

\titlerunning{A Remote-Sensing CME Analysis}

\authorrunning {S.Stamkos et al.}

\maketitle
\nolinenumbers 

\section{Introduction} \label{introduction}

Coronal Mass Ejections (CMEs) are among the most energetic and disruptive solar phenomena and release billions of tons of magnetized plasma into interplanetary space \citep{2012LRSP....9....3W}. When directed to Earth, CMEs may trigger severe geomagnetic storms, disrupt satellite functionality, pose significant radiation hazards to astronauts, and compromise critical ground-based infrastructure, including power grids and communication networks \citep{2013SpWea..11..585B}. As reliance on space-based technologies grows, understanding how CMEs evolve during heliospheric propagation is increasingly important for both heliophysics and operational space weather forecasting.

A central forecast challenge is not only to predict CME arrival times but also to anticipate geoeffectiveness, which depends strongly on the magnetic configuration at 1~AU \citep{2019RSPTA.37780096V}. Although missions such as SOHO, STEREO, and SDO have greatly improved observations of CME initiation and propagation, our ability to model the evolution of the CME internal structure during propagation remains limited. Because the orientation of the arriving field largely controls the probability of magnetopause reconnection and intense geomagnetic activity, constraining the physical processes that reshape the CME magnetic structure during transit is essential \citep{2014JGRA..119...26L}.

Once expelled from the solar corona, CMEs undergo a complex heliospheric evolution in which their early motion is set by an impulsive release of magnetic and kinetic energy, but this acceleration is brief \citep{2013SoPh..285..295V}. Solar wind drag quickly becomes the dominant influence, slowing fast eruptions toward the ambient wind speed and speeding up slow ones, while continued expansion, driven by internal magnetic and thermal pressure, carries the structure outward. Over intermediate distances, this expansion can appear approximately self-similar, yet ambient pressure gradients and structured flows introduce departures from ideal behavior, so that coherent ejecta near the Sun mature into broader, more irregular fronts by 1~AU \citep{2017SSRv..212.1159M}. In this inherently nonuniform medium, the governing forces modify the kinematics and reshape the morphology of CME, reorient the magnetic axis and alter the internal configuration \citep{2015SpWea..13..374S, 2016ApJ...823...27N}.

Several of these effects act early and can set the stage for later evolution. Deflection in the low corona can steer CMEs away from an initially radial trajectory, with the tendency controlled by coronal magnetic gradients and solar cycle conditions \citep{2015ApJ...805..168K}. Rotation can also reorient the magnetic axis over heliocentric distances of tens of solar radii, directly affecting the field orientation later encountered at 1~AU and hence geoeffectiveness \citep{2014SoPh..289.2141I}. In addition to these global geometric changes, interaction with ambient streams, coronal holes, and the large-scale heliospheric field can further distort and reconfigure the CME structure during propagation \citep{2017SSRv..212.1159M, 2018JGRA..123.7220K}.

At the leading edge, interaction with upstream plasma produces compression and a sheath region. Fast events can drive a shock, whereas slower or moderately fast events often exhibit a more gradual pile-up together with draping of magnetic field lines. The sheath itself can be geoeffective and can dominate the white-light appearance, complicating the identification of the underlying magnetic ejecta \citep{2017SSRv..212.1159M}. Simulations further indicate that pile-up can modify the apparent front and introduce asymmetries that depend on the observing geometry, complicating efforts to separate externally driven compression from intrinsic magnetic evolution using imaging alone \citep{2017SSRv..212.1159M, 2019SpWea..17.1257K}.

Beyond compression, reconnection between CME fields and the surrounding interplanetary magnetic field (IMF) has been proposed as a driver of magnetic erosion, in which portions of the CME outer flux system reconnect with the ambient field and are progressively removed during propagation \citep{2006A&A...455..349D, 2015JGRA..120...43R}. Investigating erosion observationally from remote sensing is challenging because white-light brightness primarily tracks electron density, and front-side pile-up, projection effects, and line-of-sight integration can mimic or obscure erosion-like signatures \citep{2017SSRv..212.1159M, 2010ApJ...722.1522V, 2004AnGeo..22..661O}. Forward modeling likewise shows that substantially different three-dimensional density structures can project into similar coronagraph morphologies, underscoring the nonuniqueness of single-view interpretations \citep{2006ApJ...652..763T}. These geometric ambiguities make it difficult to attribute observed structural changes to specific processes when complementary in situ magnetic field measurements are unavailable.

Magnetic erosion is the gradual loss of magnetic flux from a CME driven by magnetic reconnection with the ambient magnetic fields along the outer edge of the transient. The removal of flux alters the topology and modifies the magnetic configuration arriving at 1~AU \citep{2006A&A...455..349D, 2012JGRA..117.9101R}. In situ signatures of erosion include front-to-rear asymmetries and an imbalance in the accumulated azimuthal flux relative to idealized flux rope expectations. Additional evidence for ongoing boundary reconnection is provided by localized jets bounded by current sheets at CME interfaces \citep{2007JGRA..112.8106G}. Remote sensing diagnostics are often less direct because line-of-sight integration and Thomson-scattering geometry can preserve a smooth, symmetric white-light front even as magnetic flux is removed \citep{2006ApJ...642.1216V}.

Statistical and modeling studies suggest that erosion is common and can be substantial. A large CME sample shows frequent front and rear erosion, with average losses of about 40\% of the azimuthal flux and similar fractions of events showing front and rear signatures, based on careful boundary identification across multiple spacecraft \citep{2015JGRA..120...43R}. Numerical work highlights the role of polarity. Fronts aligned with the upstream IMF reconnect less and propagate more efficiently, whereas oppositely oriented fronts erode more strongly and can arrive later at 1~AU \citep{2019A&A...632A..89H}. Continued reconnection with distance can reduce both magnetic flux and plasma mass, narrowing and weakening the rope \citep{2021Geosc..11..314H}, and propagation models are beginning to incorporate erosion alongside drag and virtual mass to account for changes in effective cross section and momentum exchange with the solar wind \citep{2023SoPh..298...88S}. Complementary constraints of magnetic twist also suggest that highly twisted outer layers can be preferentially removed during transit \citep{2021A&A...650A.176P}. Together, these studies suggest that erosion can progressively modify the magnetic flux and plasma content of a CME during propagation. Although such changes may not strongly alter its overall appearance in white-light images, they can affect its internal magnetic structure and, consequently, its geoeffectiveness.

In situ measurements and remote sensing observations provide incomplete constraints on CME magnetic evolution. In situ instruments measure magnetic fields and plasma properties at high cadence but only along a single spacecraft trajectory, most commonly near 1~AU \citep{2006SSRv..123...31Z}. White light imaging can track CME volumes continuously across wide fields of view \citep{2012ApJ...746...64H}, but it cannot constrain the internal magnetic field and is affected by geometric projection effects that can obscure fine-scale density structure \citep{2006ApJ...652..763T, 2010ApJ...722.1522V}. Hybrid approaches that integrate modeling and observations are therefore increasingly valuable. Multi-view forward modeling, most notably the Graduated Cylindrical Shell (GCS) framework, constrains CME geometry and orientation from multi-spacecraft coronagraph data \citep{2006SPD....37.0818T, 2011ApJS..194...33T}, enabling structured comparisons between synthetic and observed density distributions as an indirect probe of evolution.

Here, we assess the potential of white-light CME observations for remotely detecting magnetic erosion. Since erosion is expected to affect the CME outer envelope by removing mass, we test whether the observed density pile-up rate is consistent with this expectation. We approach this problem by comparing detailed 3D reconstructions with synthetic images and first-order empirical descriptions of coronal density pile-up expanding on the work of \citet{2018SoPh..293...55H}. 

To test this hypothesis, we analyze a well documented CME that occurred on 27 January 2012. The event was observed simultaneously by three white light coronagraphs, COR2-A and COR2-B on the twin STEREO spacecraft, and LASCO C3 on SOHO, providing complementary perspectives that enable a more robust three-dimensional characterization of the CME as it propagates in the corona.

The manuscript is organized as follows. Section 2 describes the analyzed CME event and its geometric and kinematic properties. Section 3 details the imaging data and methods used to derive the CME front density structure, and Section 4 explains the synthetic modeling. Section 5 describes the strategy used to isolate transit effects and optimize the comparison between the observations and synthetic data. Section 6 presents the snowplow model as a theoretical description of the expected pile-up trend. Section 7 presents the results, while the remaining sections discuss them, summarize the main conclusions, and consider directions for future research.

\begin{figure*}[]
  \centering
  \includegraphics[width=16cm]{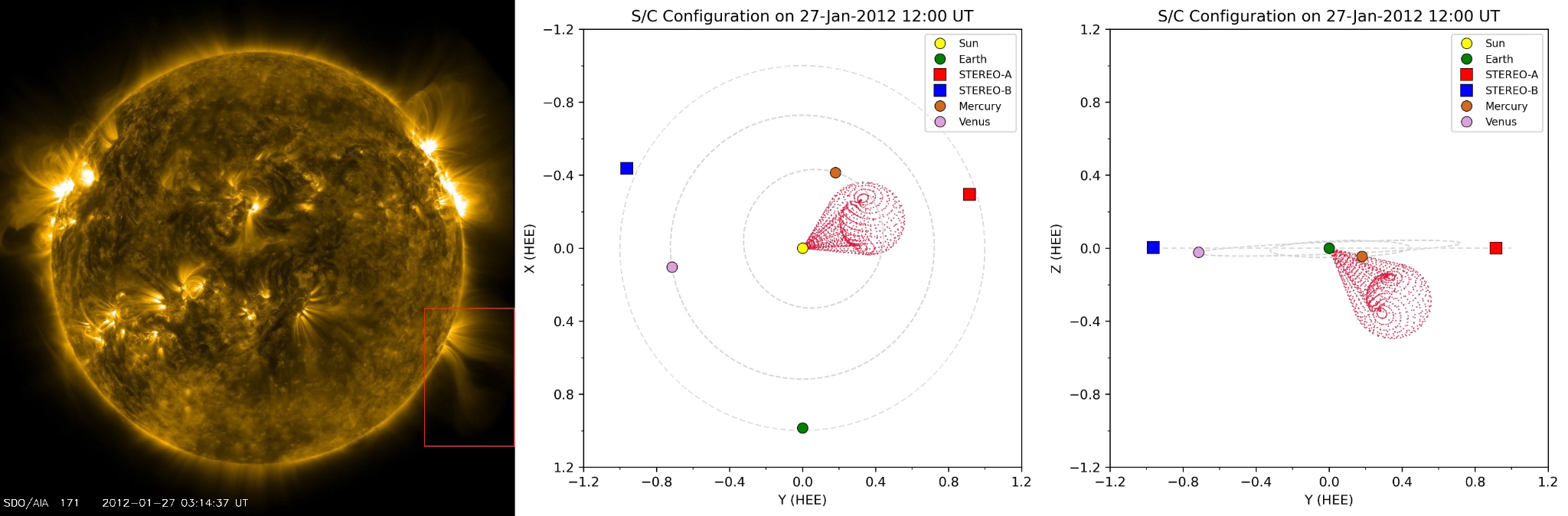}
  \caption{Overview of the 27 January 2012 CME event. Left: SDO/AIA 171\,\AA\ image at 03:14~UT capturing the early stages of the eruption near the southwestern limb. Center and right: Heliocentric Earth Ecliptic (HEE) view showing the positions of STEREO-A (red), STEREO-B (blue), and Earth or SOHO (green) on the event date. The CME orientation is indicated using the GCS wireframe, which is included here to illustrate the propagation direction and angular width.}
  \label{image1}
\end{figure*}

\section{The CME event}

\begin{figure*}[]
  \centering
  \includegraphics[width=16cm]{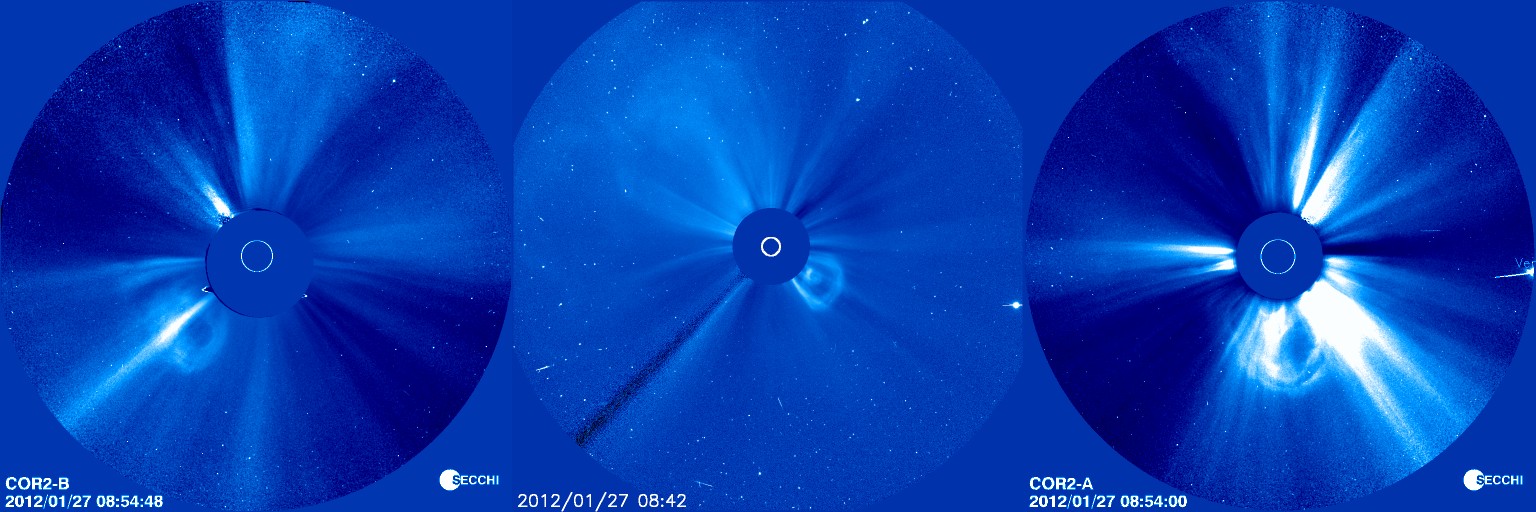}
  \caption{White-light coronagraph views of the 27 January 2012 CME from three vantage points. Left: STEREO-B/SECCHI COR2 at 08:54:48~UT. Center: SOHO/LASCO C3 at 08:42~UT. Right: STEREO-A/SECCHI COR2 at 08:54:00~UT. Panels show the CME as it propagates through the inner heliosphere.}
  \label{image2}
\end{figure*}

The CME occurred on 27 January 2012 (see Fig. \ref{image2}) and was selected for its multi-view observational coverage and well-defined morphology. The event exhibits a clear flux-rope-like morphology and first appeared in the LASCO C2 field of view at 03:47:14~UT, with a central position angle of $228^{\circ}$ in the southwest quadrant. The quantitative analysis uses STEREO/SECCHI COR2-A and COR2-B observations together with LASCO C3 data, which provide near-simultaneous viewpoints over overlapping radial ranges (COR2 $\sim2.5$--$15~R_{S}$ and C3 $\sim3.7$--$30~R_{S}$). The present study is based exclusively on remote-sensing observations and does not use event-associated in situ plasma or magnetic-field measurements.

We derived the CME kinematics by tracking its leading edge across the coronagraph observations over approximately $6.3$--$17.5~R_{S}$. At the final observation time of 12:24~UT, the quadratic fit gives an instantaneous speed of $470~\mathrm{km,s^{-1}}$, while the average speed over the tracked interval is $425~\mathrm{km,s^{-1}}$. The fit yields a modest acceleration of $0.0786~R_{S},\mathrm{hr^{-2}}\approx4.2~\mathrm{m,s^{-2}}$, with the fitted speed increasing from approximately $441~\mathrm{km,s^{-1}}$ at 10:30~UT to $470~\mathrm{km,s^{-1}}$ at 12:24~UT.

The CME remains coherent across all three viewpoints over the analyzed interval. Its leading front is well defined, and the image sequence shows no obvious evidence of substantial rotation, distortion, or deflection. Together with its near-limb source location as viewed from Earth and limited interference from neighboring structures, these characteristics make the event well suited for investigating its density evolution through multi-view observations and forward modeling.

\section{Remote-sensing observations}

The analysis uses the three coronagraph datasets introduced earlier, STEREO/SECCHI COR2-A, COR2-B, and SOHO/LASCO C3, which observed the event simultaneously. At 06:00~UT on 27 January 2012, the angular separations were approximately $107.8^{\circ}$ between STEREO-A and Earth (SOHO), and $114.4^{\circ}$ between STEREO-B and Earth (see Fig. \ref{image1}). This configuration provides complementary viewpoints and reduces line-of-sight ambiguities, allowing consistent tracking of the CME front during early propagation.

From each coronagraph sequence, we selected frames in which the leading-edge enhancement was clearly identifiable and minimally contaminated by nearby background structure. We then drew regions of interest (ROIs) to bracket this leading-edge brightness feature. Because the white-light front can include contributions from both the outermost ejecta and compressed upstream plasma, the ROIs should be interpreted as capturing the observed leading-edge density enhancement potentially part of the rear sheath, rather than as a strict ejecta-only boundary. ROIs were chosen conservatively based on contrast, spatial coherence, and consistency across viewpoints and across the image sequence.

Within each ROI, we extracted the parameters per-pixel to build a spatially resolved description of the outer structure. For each pixel we recorded the columnar electron density, the projected heliocentric distance, and the position angle. Column densities were derived from white-light brightness using Thomson-scattering inversion under a plane-of-sky assumption.

To enhance transient features relative to the static coronal background, we applied a running difference to each sequence with a two-frame offset. This highlights evolving brightness associated with the front and suppresses quasi-static background and high-frequency noise, while retaining sufficient cadence to trace the CME motion and morphology.

To derive volumetric densities from the observed surface densities we used the \citet{2018SoPh..293...55H} methodology. Each ROI is treated as the projection of a frustrum that allows  the transformation of column-integrated electron density into a volume density.

Analysis of all three coronagraphs yields a dataset that captures global expansion and local morphology, including deformations, density enhancements, and asymmetries arising from solar-wind interactions or internal evolution. Of particular interest are subtle density deviations that may signal external compression (e.g., pile-up) or internal restructuring (e.g., magnetic erosion via reconnection with the ambient magnetic field). These measurements provide the observational benchmark for the following comparative analysis.

\section{Synthetic data analysis}

We generate synthetic images of our CME using the Graduated Cylindrical Shell (GCS) framework \citep{2006ApJ...652..763T, 2009SoPh..256..111T}.  In the GCS model, the flux rope is represented as a hollow self-similarly expanding shell with a circular cross section that scales with the heliocentric distance, $a(r)=\kappa r$, with electrons concentrated on the shell rather than filling the interior.  The shell density is prescribed following \citet{2009SoPh..256..111T}, using a Gaussian profile across the shell thickness with different characteristic widths on the leading and trailing sides. Synthetic total-brightness ($tB$) images are then computed via integrating the brightness along the line of sight using Thompson scattering theory\citep{1966gtsc.book.....B, 2006ApJ...642.1216V}. 

The GCS model was applied to near-simultaneous COR2-A, COR2-B, and LASCO C3 views to derive direction, height, angular width, tilt, and aspect ratio. The GCS fitting resulted in a longitude $\approx 115^{\circ}$ and latitude $\approx -37^{\circ}$ (reported in Heliocentric Earth Ecliptic coordinates), tilt $\approx -50^{\circ}$ (defined relative to the solar equator), aspect ratio $\kappa \approx 0.36$, and half-angle $\alpha \approx 6^{\circ}$. Using those parameters, we rendered instrument-specific synthetic $tB$ images to allow pixel-wise comparisons between synthetic and observed lines of sight.

The synthetic total-brightness ($tB$) images are generated by forward modeling using a prescribed density distribution for the GCS shell. For consistency with the observational workflow, we pass the synthetic $tB$ through the same Thomson-scattering conversion operator used for the data, treating the shell as a localized enhancement near the plane of the sky. This yields an electron column-density-equivalent map in the same form and units as the observational estimates. For each ROI pixel, the conversion accounts for observer distance and viewing perspective and returns the column density together with the projected heliocentric distance and position angle. 

The resulting synthetic maps provide a spatially resolved density reference for an idealized, mass-conserving, self-similarly expanding shell. By construction, they exclude sheath pile up and coupling to the ambient solar wind. We use the synthetic maps in two complementary ways.  First, we extract synthetic profiles using the same sector definitions and sampling as the observations, enabling a controlled comparison of radial falloff rates. Second, we use the synthetic maps to quantify how much of the inferred radial behavior can be attributed to projection, sampling, and the inversion procedure when the underlying structure follows an ideal evolution. For adiabatic expansion with approximate mass conservation, the characteristic density in this reference scales as $n_e \propto r^{-3}$. After matching sampling and geometry, residual differences between the observed and synthetic profiles in slope, magnitude, symmetry, or localized structure then point to additional front evolution not represented by the ideal shell, including compression, deformation, and boundary evolution consistent with magnetic erosion.

\section{Reference-distance selection and synthetic normalization}

The observed CME front includes the effects of compression, interaction with surrounding coronal structures, and possible accumulation of plasma during propagation. The synthetic GCS shell, however, represents idealized, mass-conserving, self-similar expansion without sheath formation, swept-up plasma, or dynamical coupling to the solar wind. To compare their evolution, we place the observed and synthetic density profiles on a common scale at a reference distance, $r_{\mathrm{target}}$. This normalization preserves differences in their radial behavior while preventing an arbitrary initial density offset from being interpreted as an enhancement or deficit.

\begin{figure*}[!]
    \centering
    \includegraphics[width=\textwidth]{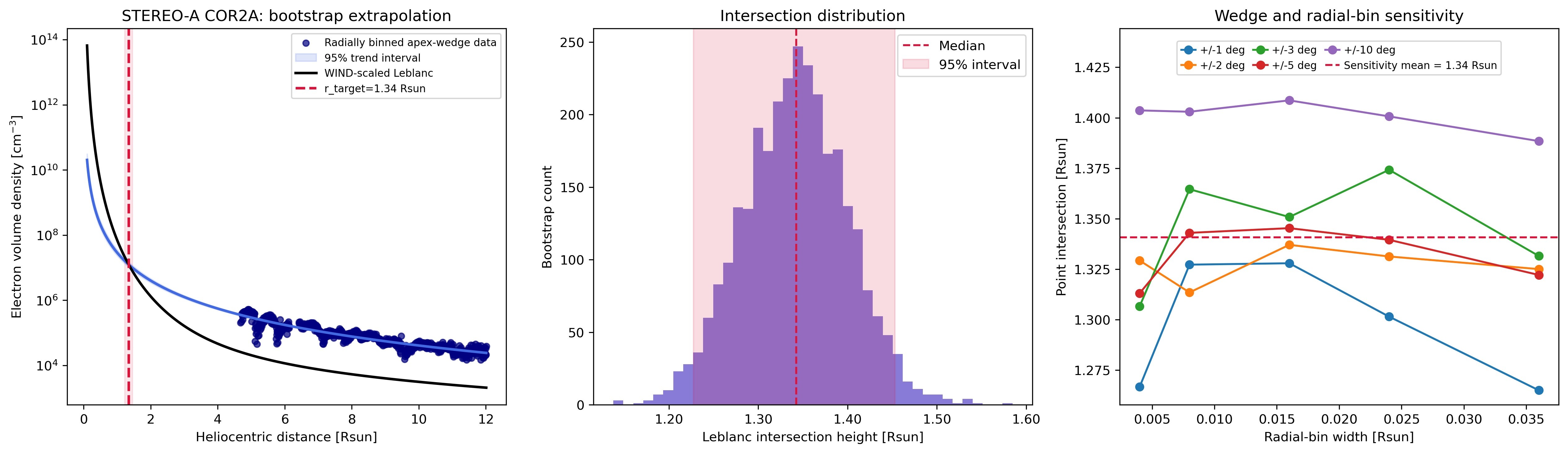}
    \caption{Reference-distance estimation for STEREO-A/COR2-A. The left panel shows the radially binned apex-wedge electron volume densities, the fitted inward extrapolation, and the WIND-normalized Leblanc density model. The middle panel shows the bootstrap distribution of the intersection distance. The right panel shows its sensitivity to the PA-wedge width and radial-bin size. The dashed red lines in the sensitivity panels indicate the mean across the tested wedge and bin configurations.}
    \label{fig:opt_data1}
\end{figure*}

We estimate this reference distance using the empirical solar-wind density model of \citet{1998SoPh..183..165L}. Since no in situ measurement is available directly along the CME propagation longitude, we normalize the background model using WIND measurements collected over a one-week interval beginning at the eruption time. The resulting mean density, $n_{\mathrm{sw}}(1~\mathrm{AU}) = 5.90 \pm 0.51~\mathrm{cm}^{-3}$, sets the amplitude of the Leblanc profile used in the comparison.

For each instrument, we initially extract the observed density profile from a wedge extending $\pm 5^{\circ}$ around the reconstructed CME apex. The electron column densities are converted to volume densities using the adopted line-of-sight depth, and the resulting radial profiles are binned, fitted, and extrapolated inward. The intersection of each extrapolated profile with the WIND-normalized background provides an estimate of the reference distance. The same procedure is applied to COR2-A, COR2-B, and LASCO C3.

To assess the sensitivity of the estimated reference distance, we vary the PA-wedge width and radial-bin size and bootstrap the densities within each bin. Each resampled profile is refitted and extrapolated to determine its intersection with the background-density model. The resulting distributions characterize the uncertainty associated with measurement scatter and profile construction. The results are shown for COR2-A in Fig.~\ref{fig:opt_data1} and for COR2-B and LASCO C3 in Fig.~\ref{fig:opt_data2}.

Across the three data sets, the extrapolated intersections range from approximately $0.87$ to $1.34~R_{\mathrm{S}}$. Guided primarily by the better-constrained COR2-A result, we adopt a common reference distance of
\begin{equation}
r_{\mathrm{target}} = 1.3~R_{\mathrm{S}}.
\label{eq:target_distance}
\end{equation}
This choice is broadly consistent with the starting-height framework of \citet{2018SoPh..293...55H}, who extrapolated CME-front density profiles to an empirical background profile to estimate the heights from which the leading-edge material may have originated. Their most frequent estimates were near $1.3 \pm 0.1~R_{\mathrm{S}}$. Here, the adopted distance serves as an extrapolated normalization reference, not as a measured onset height for expansion, compression, or magnetic erosion.

For each PA sector, the independently fitted observed and synthetic power laws are extrapolated to $r_{\mathrm{target}}$. The synthetic profile is then multiplied by the ratio of the fitted observed density to the fitted synthetic density at that distance. This adjustment makes the two fits agree at the reference point while leaving the synthetic slope and radial behavior unchanged. Their subsequent divergence therefore remains measurable. Density comparisons and reference-model calculations are defined for $r \geq r_{\mathrm{target}}$, with the departures evaluated over the observed radial range.

The normalized profiles show where the observed density exceeds or falls below the self-similar reference. Comparison with the snowplow benchmarks helps assess how much of this variation plasma accumulation can explain and where localized depletion, possibly associated with magnetic erosion, may also contribute.

\section{Snowplow benchmark for front-side pile-up}

With the self-similar shell serving as the baseline for how rapidly the density should decrease in the absence of interaction, we next estimate the amount of leading-edge compression that could flatten that trend. In heliospheric propagation, compression ahead of a CME is expected when the upstream plasma cannot fully deflect around the obstacle and instead accumulates at the front. This behavior is often described as a mass-loading process, in which the transient sweeps the slower solar-wind plasma ahead and transfers momentum to the upstream flow \citep{2006SoPh..233..233T, 2017SSRv..212.1159M}. For our density diagnostics, we treat pile-up as the accumulation of upstream solar-wind plasma at the CME leading edge and adopt a snowplow model to convert the swept-up mass into an excess density profile.

We compute this reference using a generalized one-dimensional formulation based on \citet{2018SoPh..293...55H}. In this formulation, a CME with radial speed $V_{cme}(r)$ propagating through a background wind of speed $V_{sw}(r)$ and density $N_e^{bkg}(r)$, sweeps plasma ahead of it. Assuming that the accumulated material remains confined within a thin front of radial thickness $W$ and begins to accumulate at an onset height $r_0$, the excess electron density accumulated to $r$ is

\begin{equation}
N_e(r) \;=\;
\frac{\displaystyle \int_{r_0}^{r} \left[1 - \frac{V_{sw}(r')}{V_{cme}(r')}\right]\, N_e^{bkg}(r')\, r'^2\, dr'}
     {W\, r^2 \!\left[1 - \frac{W}{r}\right]}\,.
\label{eq:snowplow}
\end{equation}

Here $r$ and $r_0$ are heliocentric distances in units of $R_{S}$, $r_0$ is the onset height used in integration, $W$ is the radial thickness of the compressed front, and $N_e(r)$ is the excess electron volume density resulting from the snowplow model. The factor $\left[1 - V_{sw}(r')/V_{cme}(r')\right]$ is retained inside the integral so that the local sweep-up efficiency varies with distance as both the CME and background wind evolve.

The background electron density $N_e^{bkg}(r)$ follows the empirical form of \citet{1998SoPh..183..165L} and is scaled by a single multiplicative factor fixed at $1~\mathrm{AU}$. The background flow speed $V_{sw}(r)$ is prescribed as a smooth rise from low-coronal values to an asymptotic wind speed, implemented with a tanh-like transition characterized by an asymptote $V_{\infty}$ and two shape parameters that set the location and width of the acceleration region \citep{2012PhDT.......120K}. Since Eq.~\eqref{eq:snowplow} depends on the ratio $V_{sw}/V_{cme}$, this prescription allows the sweep-up efficiency to vary radially rather than imposing a constant solar-wind speed.

CME kinematics are derived from the three-dimensional reconstruction performed with the PyThea forward-modeling code \citep{2022FrASS...9.4137K}. For the event studied here, we extract the apex coordinates of the reconstructed front over time, build the height--time curve, differentiate to obtain $V_{cme}(t)$, and fit a low-order polynomial in $r$ to produce a smooth $V_{cme}(r)$ for use in Eq.~\eqref{eq:snowplow}. 

The remaining geometric parameter is the radial shell thickness, $W$. For each position-angle sector, we estimate $W$ from the coronagraph data using the same front ROIs employed for the density extraction. Specifically, we measure the radial extent of the front, defined as the outer minus inner ROI boundary, in each frame, and then take the mean over the image sequence. This yields a single thickness, $W_{\mathrm{eff}}(\phi)$, for each PA sector, which we adopt as $W$ in Eq.~\eqref{eq:snowplow} for that sector. For the PA interval analyzed here, $157^\circ$--$197^\circ$, the individual measurements span $W_{\mathrm{eff}}\simeq 0.33$--$2.31,R_{S}$ over heliocentric distances of $r\simeq 4.41$--$11.61,R_{S}$. The larger values generally occur at greater distances, reflecting the increasing radial extent of the front within the ROI. The PA-sector means used in the snowplow calculation range from $0.802$ to $1.195,R_{S}$. Combining the sector means with inverse-variance weighting gives a representative value of $\langle W_{\mathrm{eff}}\rangle \simeq 0.95 \pm 0.03,R_{S}$. Using a sector-specific piecewise constant $W_{\mathrm{eff}}(\phi)$ preserves the analytic form of the snowplow model while allowing the calculation to reflect the thickness of the observed front in each PA bin.

\section{Results}

\begin{figure}[t]
    \centering
    \includegraphics[width=\columnwidth]{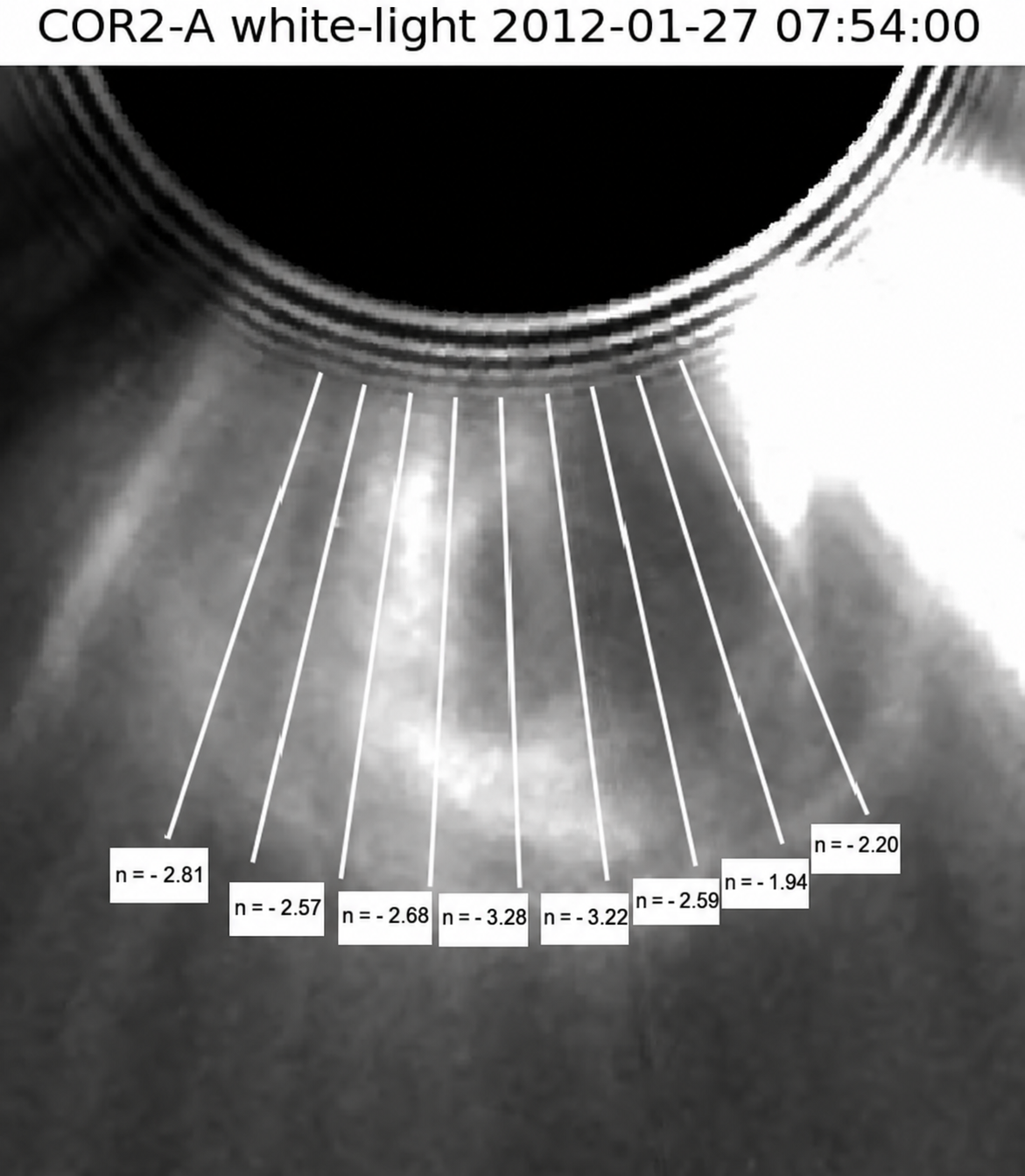}
    \caption{Example of the COR2-A sector masks overlaid on a representative frame. The analyzed front is divided into eight $5^{\circ}$ PA sectors. Selected labels indicate the observed power-law exponents obtained from the corresponding radial profiles in Fig.~\ref{fig:cor2a_sectors}.}
    \label{fig:pa_sectors}
\end{figure}

\begin{table*}[!]
    \centering
    \caption{Fitted density indices across grouped PA intervals.}
    \label{tab:pa_density_indices}
    \begin{tabular}{llccl}
        \toprule
        Instrument & PA range ($^{\circ}$)
        & $n_{\mathrm{obs}}$ & $n_{\mathrm{syn}}$
        & Observed trend \\
        \midrule
        COR2-A
        & 157--172 & $-3.0$ to $-2.1$ & $-2.7$ to $-2.5$ & Mixed \\
        & 172--182 & $-3.4$ to $-3.1$ & $-2.8$ to $-2.6$ & Steeper \\
        & 182--187 & $-2.9$ to $-2.5$ & $-2.8$ to $-2.5$ & Close to synthetic \\
        & 187--197 & $-2.4$ to $-1.7$ & $-2.8$ to $-2.6$ & Flatter \\
        \addlinespace
        COR2-B
        & 119--131 & $-2.7$ to $-2.3$ & $-2.8$ to $-2.5$ & Mixed, often flatter \\
        & 131--137 & $-3.4$ to $-2.9$ & $-2.7$ to $-2.5$ & Steeper \\
        & 137--146 & $-3.0$ to $-2.7$ & $-2.8$ to $-2.4$ & Generally steeper \\
        & 146--151 & $-3.2$ to $-2.8$ & $-2.9$ to $-2.6$ & Generally steeper \\
        \addlinespace
        LASCO C3
        & 221--231 & $-3.3$ to $-2.3$ & $-2.9$ to $-2.5$ & Mixed, mostly steeper \\
        & 231--245 & $-3.3$ to $-2.8$ & $-3.0$ to $-2.4$ & Generally steeper \\
        & 245--254 & $-3.8$ to $-2.7$ & $-2.8$ to $-2.4$ & Steepens toward higher PAs \\
        \bottomrule
    \end{tabular}

    \tablefoot{The density index $n$ is defined by $n_e(r)\propto r^{~n}$. Each range summarizes the constituent sectors rather than their fitting uncertainties. Trend descriptions refer to the fitted central values relative to the corresponding synthetic profiles.}
\end{table*}

\begin{figure*}[!]
  \centering
  \includegraphics[width=19cm]{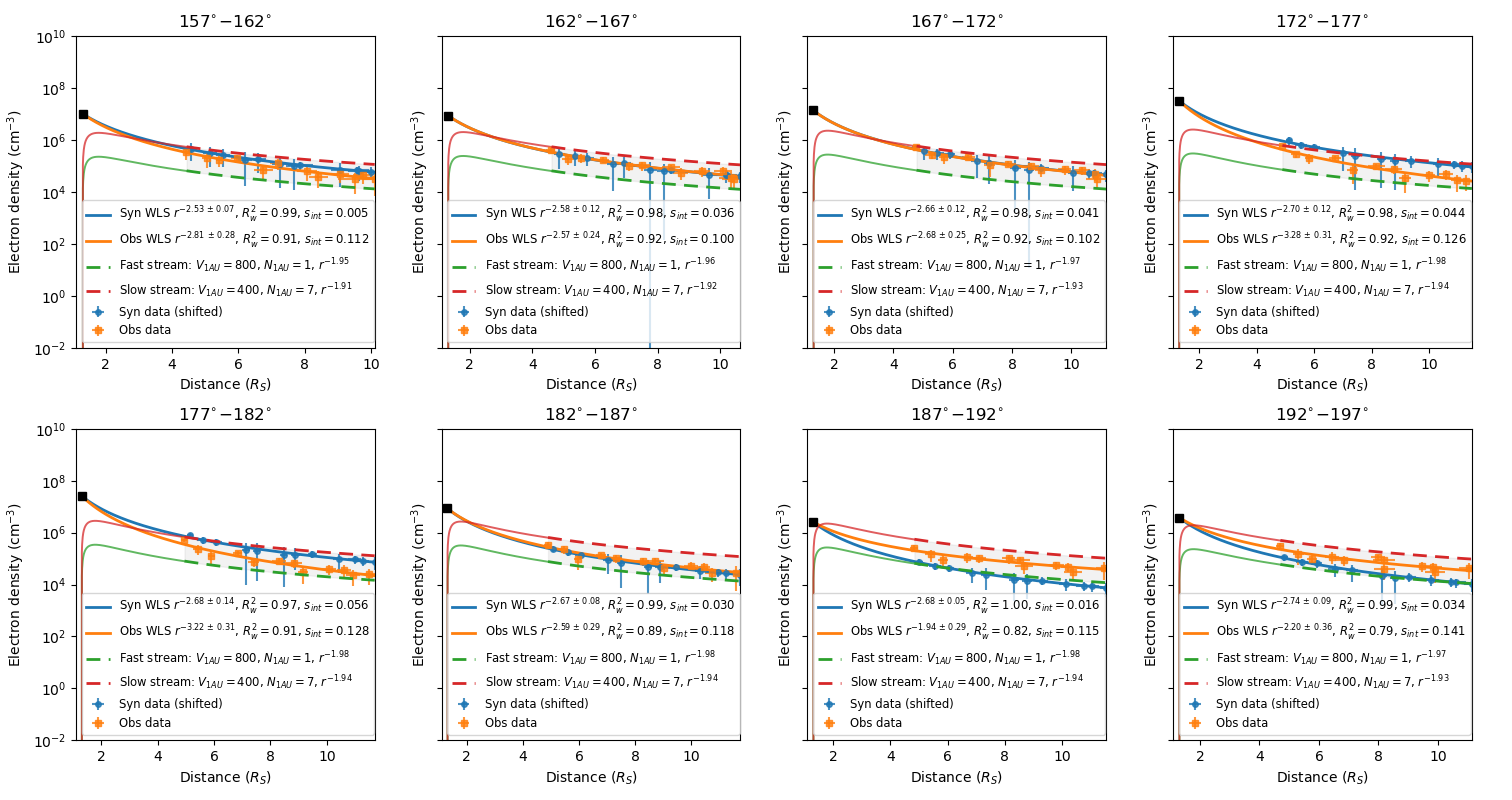}
    \caption{Sector-wise density comparison for COR2-A. Orange points show the observations, and blue points show the synthetic profiles normalized at $r_{\mathrm{target}}=1.3~R_{\mathrm{S}}$, marked by the black square. Error bars are based on the within-sector pixel standard deviations. Solid curves show weighted least-squares power-law fits, with the fitted exponents, $R_w^2$, and intrinsic scatter $s_{\mathrm{int}}$ reported in each panel. The fast- and slow-stream snowplow curves provide compression benchmarks for the adopted upstream conditions. Their representative power-law slopes are obtained by least squares over the radial span covered by the density profiles.}
  \label{fig:cor2a_sectors}
\end{figure*}

Using the comparison reference established in the previous section, we examine how the observed density evolution varies across the CME front and departs from the synthetic self-similar shell. COR2-A provides the clearest coverage for this event and is therefore the primary diagnostic view. The same procedure is applied to COR2-B and LASCO C3, with their detailed results presented in Appendices~\ref{Cor2b_results} and~\ref{LASCO-C3_results}.

The front is divided into narrow position-angle sectors, as illustrated in Fig.~\ref{fig:pa_sectors}. The analysis concentrates on the region around the reconstructed apex, where the leading front can be sampled most consistently. Toward the flanks, a PA sector can intersect a broader radial extent and include material from the inner ejecta. Projection and line-of-sight overlap can therefore make the measured width less representative of the leading-front thickness.

At each time step, we calculate the mean electron density and mean heliocentric distance of the pixels within each sector's ROI mask. For COR2-A, a $5^{\circ}$ sector contains approximately $7300$ pixels on average before outlier rejection, which removes approximately $4\%$--$5\%$ per sector and frame. The retained density distributions are approximately Gaussian. Their sample standard deviation describes the spatial variability within the sector and determines the plotted density spread. The formal standard error of the mean, obtained by dividing this standard deviation by the square root of the retained pixel count, is used to construct the initial fitting weights.

We assess the sensitivity to angular resolution by repeating the analysis with sector widths from approximately $1^{\circ}$ to $10^{\circ}$. As expected, narrower sectors increase scatter and uncertainty because fewer pixels contribute to each sector mean, and the results become slightly more sensitive to small-scale structure and ROI-edge effects. Broader sectors reduce statistical noise but can progressively wash out the small propagation effects we aim to capture. The adopted $5^{\circ}$ binning therefore represents a compromise: it samples a substantial portion of the front without misrepresenting the structure, while maintaining sufficient pixel statistics for stable weighted fits.

For each sector, we fit power laws of the form $n_e(r)\propto r^{~n}$ using weighted least squares (WLS). The synthetic profiles are normalized at $r_{\mathrm{target}}=1.3~R_{\mathrm{S}}$ following the procedure described above. This multiplicative adjustment changes their amplitudes without changing their fitted indices. We report $n$, its formal uncertainty, and the weighted coefficient of determination, $R_w^2$.

The scatter around a fitted power law can exceed that expected from the formal standard errors. This additional dispersion may reflect unresolved density structure, residual background-subtraction or modeling effects, and local departures from a single power-law trend. We account for it using an intrinsic-scatter term, $s_{\mathrm{int}}$, which represents variation not captured by the formal uncertainties. It is combined in quadrature with the uncertainties used in the fit:

\begin{equation}
\sigma_{\mathrm{eff},i}^{2}
=
\sigma_i^{2}
+
s_{\mathrm{int}}^{2},
\label{eq:density_intrinsic_scatter}
\end{equation}

where $\sigma_i$ denotes the formal uncertainty used in the regression. When the initial fit gives $\chi_{\nu}^{2}>1$, we adjust $s_{\mathrm{int}}$ and refit until $\chi_{\nu}^{2}\simeq1$. Otherwise, $s_{\mathrm{int}}=0$. This procedure leaves the measured densities unchanged while preventing points with very small formal uncertainties from dominating the fit. The uncertainty on the fitted index then reflects the additional dispersion, while $R_w^2$ describes the agreement between the power law and the data under the adjusted weights.

\begin{figure}[]
  \centering
  \includegraphics[width=\columnwidth]{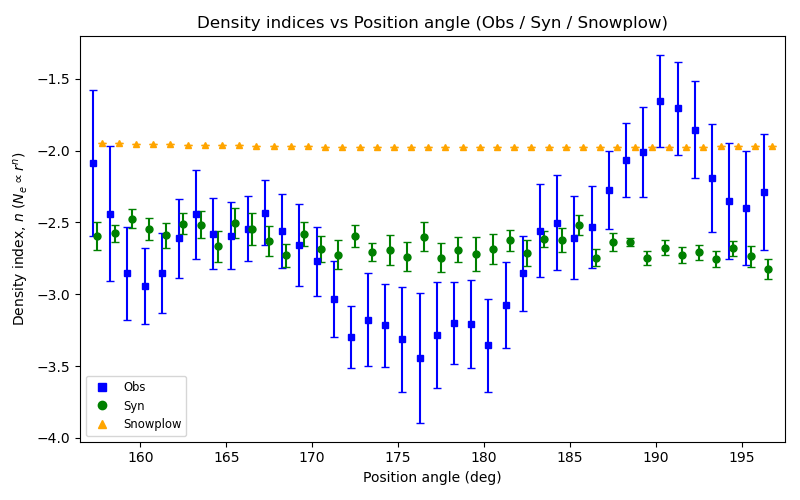} 
  \caption{Radial density fall-off indices $n$ versus position angle for the observations (blue squares), the synthetic CME (green circles), and the snowplow model (orange triangles). Each point corresponds to a PA bin of $1^\circ$ and vertical bars show uncertainties on $n$. Flattening toward the snowplow value marks compression-dominated sectors, while indices more negative than the synthetic shell identify sectors with steeper-than-self-similar falloff.}
  \label{fig:density_index_vs_pa}
\end{figure}

Figure~\ref{fig:cor2a_sectors} presents the COR2-A density profiles and snowplow benchmarks (\ref{snowplow_benchmarks}). The finer PA dependence of the fitted density indices is shown in Fig.~\ref{fig:density_index_vs_pa}, while Table~\ref{tab:pa_density_indices} summarizes their approximate ranges across all three viewpoints.

In COR2-A, the central sectors at $172^{\circ}$--$182^{\circ}$ show the clearest steepening relative to the synthetic shell, while the higher-PA sectors at $187^{\circ}$--$197^{\circ}$ show pronounced flattening toward the snowplow-like slope. The lower-PA sectors display mixed behavior, and the profiles approach the synthetic trend again around $182^{\circ}$--$187^{\circ}$. COR2-B also shows mixed behavior at lower PAs, followed by predominantly steeper observed profiles. LASCO C3 exhibits a similarly variable pattern, with the strongest steepening toward its highest PAs. These differences are summarized in Table~\ref{tab:pa_density_indices}.

The three viewpoints therefore reveal a spatially structured departure from self-similar evolution, although their detailed PA patterns differ. Their sectors sample different projected portions of the front over different radial intervals, so a one-to-one angular correspondence is not expected. The COR2-A flattening is consistent with localized accumulation, while the steeper profiles identify regions where density decreases more rapidly than in the synthetic reference. This contrast motivates the subsequent density-excess and snowplow comparisons, which assess the contributions of density accumulation, depletion possible redistribution.

\subsection{Pile-up quantification}

To quantify the excess density ahead of the front at each moment, we process the COR2-A sequence frame by frame. At every timestamp we sample the front in eight narrow position-angle (PA) sectors per frame.

For each contributing sector $k$ at timestamp $i$, we have an observed electron density $n_{obs}^{(k,i)}$ and a synthetic expectation $n_{syn}^{(k,i)}$, both referenced to the same level used in the fits. A sector is tagged as pile-up only when the observation lies unambiguously above the synthetic after uncertainties are included,

\begin{equation}
\bigl(n_{obs}^{(k,i)} - \varepsilon_{obs}^{(k,i)}\bigr)
\;>\;
\bigl(n_{syn,sh}^{(k,i)} + \varepsilon_{syn}^{(k,i)}\bigr),
\label{eq:nonoverlap}
\end{equation}

where $n_{syn,sh}$ is the vertically shifted synthetic profile used for plotting and fitting, and $\varepsilon_{obs}$ and $\varepsilon_{syn}$ are the per-point uncertainty proxies. If the error bars touch, the sector is not counted as pile-up.

For every sector and timestamp we define the non-negative absolute excess
\begin{equation}
e^{(k,i)} \;=\;
\max ~\!\Bigl(n_{obs}^{(k,i)} - n_{syn,sh}^{(k,i)},\, 0\Bigr),
\label{eq:excess}
\end{equation}
and an indicator $C_{PU}^{(k,i)}$ that equals one only when the non-overlap test is satisfied \eqref{eq:nonoverlap}. With $N$ contributing sectors in image $i$, we summarize the front-wide excess in two ways,
\begin{equation}
\overline{e}_i \;=\; \frac{1}{N}\sum_{k=1}^{N} C_{PU}^{(k,i)}\, e^{(k,i)},
\qquad
f_i \;=\;
\frac{\displaystyle \sum_{k=1}^{N} C_{PU}^{(k,i)}\, e^{(k,i)}}
     {\displaystyle \sum_{k=1}^{N} n_{obs}^{(k,i)}} \times 100\% .
\label{eq:frontwide}
\end{equation}
Here $\overline{e}_i$ is the mean absolute excess across sectors at that timestamp, and $f_i$ is the fraction of the total observed density at that timestamp attributable to pile-up. Because $f_i$ is normalized by the sector-summed $n_{obs}$, sectors with larger density contribute more to the share, while $\overline{e}_i$ is a simple average that assigns zero to non–pile-up sectors.

\begin{table}[t]
\centering
\caption{Clear pile-up detections in COR2-A.}
\label{tab:cor2a_pileup_summary}
\begin{tabular}{cccc}
\hline
Time (UT) & Pile-up sectors & $f_i$ (\%) &
$\overline{e}_i$ ($10^{3}\,\mathrm{cm^{-3}}$) \\
\hline
07:39 & $2/8$ & 11.9 & 49.68 \\
07:54 & $1/8$ & 5.5  & 11.72 \\
08:39 & $1/8$ & 6.3  & 10.59 \\
08:54 & $1/8$ & 11.1 & 10.12 \\
09:24 & $2/8$ & 21.3 & 22.36 \\
09:39 & $2/8$ & 22.9 & 18.60 \\
09:54 & $1/8$ & 9.9  & 4.489 \\
10:24 & $2/8$ & 19.0 & 10.41 \\
10:39 & $2/8$ & 17.8 & 9.227 \\
10:54 & $1/8$ & 8.9  & 2.889 \\
11:24 & $2/8$ & 25.7 & 8.411 \\
\hline
\end{tabular}
\tablefoot{$f_i$ is the front-wide fraction identified as clear pile-up at each timestamp, and $\overline{e}_i$ is the corresponding mean excess electron density.}
\end{table}

Taken together, these measurements show that clear pile-up is detected in many, but not all, COR2-A frames and is not uniformly distributed across the front. In the timestamps listed in Table~\ref{tab:cor2a_pileup_summary}, the number of sectors satisfying the conservative non-overlap criterion ranges from $0/8$ to $2/8$. The pile-up share of the index density ranges from $0.0\%$ to $25.7\%$, while the mean absolute excess density ranges from $\overline{e}_i \simeq 0.000\times10^{0}$ to $4.968\times10^{4}\ \mathrm{cm^{-3}}$.

The first listed frame shows the largest absolute excess, with clear pile-up in $2/8$ sectors and $\overline{e}_i \simeq 4.968\times10^{4}\ \mathrm{cm^{-3}}$. One frame, at 08:24~UT, shows no sector satisfying the conservative pile-up criterion. At the remaining times, the detected excess is confined to only $1/8$ or $2/8$ sectors, but the pile-up share of the index density remains non-negligible. The largest fractional contribution occurs at 11:24~UT, where the pile-up share reaches $25.7\%$, even though clear pile-up is detected in only $2/8$ sectors.

Because a sector is counted only when its lower observational bound exceeds the upper synthetic bound, these fractions are conservative by construction. They should therefore be interpreted as robust detections of clear excess density, not as a complete estimate of all compressed plasma along the front. Combined with the sector-wise slope analysis, the time-resolved pile-up measurements support a front in which compression is significant, intermittent in time, and concentrated in selected PA sectors rather than distributed uniformly across the CME front.

\subsection{Application of the snowplow benchmarks to the CME front}
\label{snowplow_benchmarks}

We evaluate Eq.~\eqref{eq:snowplow} from the empirically chosen reference distance $r_0$ to the maximum heliocentric distance covered by each sector. To bracket plausible upstream conditions, we compute two snowplow realizations using fast-stream and slow-stream backgrounds normalized at $1~\mathrm{AU}$ ($\simeq 215~R_{\mathrm{S}}$). The fast-stream case provides a lower pile-up benchmark, while the slow-stream case provides an upper pile-up benchmark within the adopted one-dimensional assumptions. We then compare, sector by sector, the LOS-corrected observations, the self-similar synthetic shell, and the two snowplow benchmarks.

For each sector, we fit a power law $n_e(r)\propto r^{~n}$ over the sampled radial range and report $n_{\mathrm{snow}}$ alongside $n_{\mathrm{obs}}$ and $n_{\mathrm{syn}}$, following the index convention introduced above. Over the COR2-A range, the snowplow solutions follow an approximately $r^{-2}$ trend, corresponding to $n_{\mathrm{snow}}\simeq-2$ and providing a practical compression reference.

Figure~\ref{fig:density_index_vs_pa} shows the radial density index $n$ as a function of position angle for the observations, the synthetic shell, and the snowplow benchmarks. The COR2-A results show mixed behavior across PA. From the eastern side through the central front, several observed indices exceed the synthetic indices, with the steepest sectors reaching approximately $n\simeq-3.4$. Toward the western side, the observed indices decrease toward $n\simeq-2$ in the highest-PA sectors. These western sectors are therefore closer to the snowplow benchmark and are consistent with enhanced compression.

The synthetic indices remain comparatively stable across PA, while the observed indices show stronger sector-to-sector variability. Sectors where $n_{\mathrm{obs}}>n_{\mathrm{syn}}$ exhibit flattening relative to the synthetic shell, consistent with a contribution from plasma accumulation. In contrast, sectors where $n_{\mathrm{obs}}<n_{\mathrm{syn}}$ show a steeper density decline than expected from the idealized shell. These steeper profiles are not reproduced by the approximately $r^{-2}$ snowplow benchmark and identify candidate regions where depletion, redistribution of material, or erosion-like processes may contribute.

The complementary COR2-B and LASCO C3 results reinforce this interpretation. They do not reproduce a simple one-to-one PA trend because each viewpoint samples a different projection of the front, but both show departures of the observed indices from the synthetic shell. In COR2-B, several higher-PA sectors are markedly steeper than the synthetic reference. In LASCO C3, some sectors remain near the synthetic trend, while others become steeper. Together, the three viewpoints reveal a spatially variable density evolution, spanning sectors that approach the snowplow-like compression trend and sectors that decline more steeply than the synthetic reference.

The snowplow benchmarks therefore help identify where accumulation can account for the observed evolution and where additional processes need to be considered.

\subsection{Density deficit relative to the snowplow prediction}

To quantify the degree to which the observed density front falls below the snowplow expectation, we define, for each PA sector and heliocentric distance $r$, the ratio
\begin{equation}
\mathcal{R}(r,\mathrm{PA}) =
\frac{N_{e,\mathrm{obs}}(r,\mathrm{PA})}
     {N_{e,\mathrm{snow}}(r,\mathrm{PA})},
\end{equation}
where $N_{e,\mathrm{obs}}$ is the observed electron density and $N_{e,\mathrm{snow}}$ is the density predicted by the snowplow model evaluated at the same height. From this ratio we define the density deficit percentage as
\begin{equation}
D(r,\mathrm{PA}) =
\left[
1 -
\frac{N_{e,\mathrm{obs}}(r,\mathrm{PA})}
     {N_{e,\mathrm{snow}}(r,\mathrm{PA})}
\right]\times 100\%.
\end{equation}

With this definition, $D=0\%$ corresponds to exact agreement between the observations and the snowplow prediction. Positive values indicate that the observed front contains less material than expected from the model, while negative values indicate locations where the observed density exceeds the snowplow prediction.

We evaluate the density deficit for two snowplow realizations representing slow- and fast-stream conditions. The calculation is performed over eight $5^{\circ}$-wide PA sectors spanning $157.0^{\circ}$--$197.0^{\circ}$. For each realization, we calculate $D(r,\mathrm{PA})$ at every sampled distance and summarize the results using sector-averaged values. The sign of the inferred deficit depends strongly on the adopted snowplow background. In the fast-stream realization, the observed density exceeds the snowplow prediction at every sampled point, producing negative values of $D(r,\mathrm{PA})$. These negative values represent a density excess relative to the fast-stream benchmark rather than a density deficit.

The slow-stream realization produces the opposite result. The observed density remains below the snowplow prediction at every sampled point and in every PA sector, resulting in positive density deficits throughout the analyzed region. The mean sector deficits range from approximately $47.2\%$ to $69.1\%$. The weakest mean deficit occurs in the $167.0^{\circ}$--$172.0^{\circ}$ sector, where it is approximately $47.2\%$, while the strongest occurs in the $177.0^{\circ}$--$182.0^{\circ}$ sector, where it reaches approximately $69.1\%$.
The slow-stream realization also exhibits a clear sector-to-sector variation. From $157.0^{\circ}$ to $177.0^{\circ}$, the mean deficits are approximately $59.2\%$, $52.9\%$, $47.2\%$, and $59.6\%$. The largest deficits occur in the central sectors, reaching approximately $69.1\%$ and $67.4\%$. Toward the higher-PA flank, the mean deficit remains substantial but becomes somewhat weaker, decreasing to approximately $62.4\%$ and $58.8\%$.

The point-by-point radial evolution is nonmonotonic and is therefore more appropriately interpreted in terms of broad sector-scale tendencies than as a continuous increase with distance. In several sectors, the largest deficits occur at greater heliocentric distances. For example, in the sector with the largest mean deficit, $D$ is approximately $26.1\%$ at the innermost measurement near $4.92~R_S$ but reaches approximately $75\%$--$86\%$ at several larger distances, despite fluctuations among the intermediate points. A similar behavior is found in the neighboring sector, where the deficit is approximately $45.8\%$ near $4.88~R_S$ and reaches approximately $76\%$--$81\%$ at selected larger distances.

Overall, the comparison demonstrates that the inferred density deficit is strongly dependent on the adopted snowplow background. The fast-stream realization underpredicts the observed front density and therefore yields a density excess relative to that benchmark. In contrast, the slow-stream realization overpredicts the observed density throughout the analyzed PA range, producing mean deficits of approximately $47\%$--$69\%$. Within the two realizations considered here, positive density deficits are therefore obtained only relative to the slow-stream benchmark, with the strongest sector-averaged deficits occurring near the central portion of the analyzed front. The deficit should consequently be interpreted as a background-relative quantity rather than as a model-independent measurement of density depletion.

\section{Discussion}

The sector-wise density slopes show that the CME front did not follow a uniform density-evolution trend. In COR2-A, the higher-PA western sectors display the clearest flattening relative to the self-similar synthetic shell, with the strongest case approaching snowplow-like behavior. The lower and central sectors either remain close to the synthetic reference or become locally steeper, which is inconsistent with uniformly distributed pile-up. COR2-B (Fig.~\ref{fig:cor2b_sectors}) and LASCO C3 (Fig.~\ref{fig:c3_sectors}) do not reproduce the COR2-A PA pattern, but each observes a different projection of the three-dimensional front. Line-of-sight integration and Thomson-scattering geometry can mix different structures within the sector averages, while instrumental response also affects their visibility. The catalog analysis of \citet{2017ApJ...838..141V} found that COR2-A recorded more CMEs than COR2-B and that non-detection became more frequent for events farther from the plane of the sky. The authors related the lower COR2-B detection rate partly to its greater stray-light background and lower signal-to-noise ratio. A subsequent visibility analysis confirmed its lower CME detectability relative to COR2-A and LASCO \citep{2020ApJ...900..161V}. These statistics do not explain the fitted slopes directly, but they provide context for the cleaner sector separation in COR2-A. We therefore retain COR2-A as the primary diagnostic viewpoint, with COR2-B and C3 providing complementary checks across viewing geometries and heliocentric distances.

Pile-up provides a natural explanation for flattened density profiles because solar-wind plasma accumulated ahead of the CME increases the electron column density and Thomson-scattered brightness relative to a freely expanding shell. The snowplow calculation supplies two transparent reference cases: a fast, tenuous stream with relatively weak accumulation and a slower, denser stream with stronger accumulation. Sampling the Predictive Science Inc. MAS global MHD solution along the event direction gives $V_{\rm sw}\sim780~\mathrm{km\,s^{-1}}$ and a proton density of order $1~\mathrm{cm^{-3}}$ at $1~\mathrm{AU}$, supporting the fast-stream realization \citep{1999PhPl....6.2217M,2011SoPh..274..361R}. However, the MAS solution is model-dependent and cannot uniquely establish the conditions encountered by the CME, so both realizations are retained as sensitivity bounds. Any inferred excess or deficit is consequently conditional on the chosen upstream background.

The timestamp-resolved pile-up classification indicates excess density in only a subset of sectors and times. Because the criterion requires non-overlapping uncertainty bounds, it deliberately selects only well-separated cases and supports spatially localized rather than uniform compression. Sector averaging can nevertheless dilute the compressed patches and neighboring depleted regions. Conversely, the steeper-than-synthetic sectors are compatible with reduced accumulation, redistribution, or loss of material, but density slopes alone cannot establish magnetic erosion because they are also influenced by projection, front curvature, Thomson-scattering geometry, and spatial sampling. Reconnection exhausts observed at ICME boundaries and multipoint evidence for patchy erosion within the inner heliosphere establish that localized erosion is physically plausible \citep{2007JGRA..112.8106G,2012JGRA..117.9101R}. The steepened sectors therefore identify erosion-sensitive regions for follow-up rather than direct detections of reconnection.

Internal flux-rope restructuring offers an additional interpretation. \citet{2014PhRvL.113c1101F} considered an ICME elongated in the latitudinal direction during propagation and showed analytically that this deformation can drive the flux rope away from a minimum-energy Taylor state. Their MHD simulations indicated that relaxation may then proceed through reconnection at multiple internal sites, producing stochastic fields and reorganizing the original flux-rope structure. If this relaxation begins within the coronagraph field of view, it could redistribute plasma and magnetic flux before the CME reaches $1~\mathrm{AU}$. That study did not predict a white-light signature, so the connection remains tentative, and the present density measurements cannot distinguish internal restructuring from erosion at the CME--solar-wind boundary. Both processes may contribute to sector-dependent depletion.

The range of observed behavior also reflects limitations of the one-dimensional snowplow representation. It cannot describe lateral flow around the CME, curvature-dependent accumulation, or redistribution along the front. Moreover, the upstream medium is structured by streamers, coronal holes, and the heliospheric magnetic field, while preconditioning by earlier transients may modify the local plasma environment. These effects can cause the efficiency of pile-up and the radial density falloff to vary across the same CME front.

The density-deficit proxy quantifies the dependence on the assumed background through $\mathcal{R}(r,\mathrm{PA})=N_{e,\mathrm{obs}}/N_{e,\mathrm{snow}}$ and $D(r,\mathrm{PA})=[1-\mathcal{R}(r,\mathrm{PA})]\times100\%$. Relative to the fast-stream realization, the observations are generally denser than predicted and $D$ is negative, showing that this benchmark underestimates the front density. Relative to the slow-stream realization, $D$ is positive in every PA sector, with sector-averaged values of approximately $47\%$--$69\%$ and the largest deficits near the central sectors. Thus, $D$ is a benchmark-dependent density diagnostic rather than a direct measurement of magnetic erosion. Positive deficits may indicate that swept-up material is not fully retained within the observed compressed region, that the front broadens or disperses, that the one-dimensional approximation is inadequate, or that external or internal erosion-related processes contribute to depletion. The proxy alone cannot discriminate among these possibilities.

The dependence of the inferred deficit on the upstream background extends beyond the fast- and slow-stream cases. Both calculations use the Leblanc radial density profile, so they test different wind conditions within the same prescribed density law. A different radial profile, even if scaled to the same density at $1~\mathrm{AU}$, would change the plasma available for accumulation and hence the snowplow density and inferred deficit. Determining whether another profile is more appropriate requires independent, event-specific constraints on the density and flow along the CME trajectory. Such constraints would provide a stronger basis for interpreting the deficit.

This interpretation extends the white-light pile-up analysis of \citet{2018SoPh..293...55H}. They tracked a narrow $10^{\circ}$ region at the CME front through the LASCO C2 and C3 fields of view and converted the line-of-sight signal to a volume electron density using a frustum volume and a plane-of-sky assumption. For 13 three-part CMEs, they measured front densities close to an $r^{-3}$ falloff over approximately $2$--$30~R_{S}$ and found no detectable pile-up. Their study also introduced the one-dimensional snowplow formulation used here as a benchmark. Snowplow is only one member of a wider family of compression treatments, which includes analytic swept-mass approximations, global MHD simulations, shock and stand-off relations, and propagation models with added-mass effects. Its purpose here is not to reproduce the complete CME front, but to provide transparent bounding cases that can be applied consistently across sectors. Our multi-viewpoint observations, GCS-based synthetic shell, and azimuthally resolved profiles then allow each sector to be compared with both self-similar and snowplow expectations. Taken together, the slope analysis, pile-up classification, and the deficit proxy show that some sectors are consistent with compression, others remain close to self-similar expansion, and others exhibit depletion consistent with external erosion, internal restructuring, or both.

\section{Conclusions}

In this work, we developed and tested a remote-sensing methodology for diagnosing CME-front evolution during propagation using density-sensitive white-light observations. The approach combines simultaneous multi-view coronagraph data with two complementary references: an idealized, self-similar GCS-based shell that provides a baseline for unperturbed expansion after matching the observational sampling and viewing geometry, and an analytic snowplow treatment that describes compression-driven pile-up from swept-up background plasma.

Applied to the CME studied here, the sector-wise power-law fits reveal that the density evolution is not uniform around the front. In COR2-A, the western sectors show the clearest flattening of the observed profiles, with slopes approaching the snowplow-like compression benchmark. At lower and central PAs, several observed profiles decline more steeply than the synthetic shell. These sectors cannot be accounted for by compression alone and therefore represent the strongest candidates for density depletion or erosion-like evolution within the present framework.

COR2-B and LASCO C3 support the broader conclusion that CME-front density evolution is spatially structured, although the detailed PA dependence varies among viewpoints. This variation is expected because each instrument samples a different projection of the three-dimensional front under different visibility conditions. The multi-view comparison therefore strengthens the interpretation of sector-dependent evolution while emphasizing that individual density slopes must be considered in the context of projection, sampling, and instrumental sensitivity.

The snowplow comparison further demonstrates the importance of the assumed upstream conditions. The observed front density is generally higher than the fast-stream prediction, so this realization does not produce a density deficit. In contrast, the slow-stream realization overpredicts the observed density across all analyzed PA sectors, resulting in positive deficits relative to that benchmark. The inferred degree of pile-up or density deficit therefore depends strongly on the adopted background wind, and the observed sector structure cannot be fully represented by a single one-dimensional accumulation model.

Taken together, the observed flattening and steepening indicate a mixed front in which localized pile-up coexists with regions of reduced density retention. Flattening relative to the synthetic shell is consistent with accumulation, whereas steepening identifies sectors where density depletion, redistribution of material, external erosion, or internal restructuring may contribute. Because density is an indirect tracer of magnetic erosion, these signatures do not constitute standalone evidence of reconnection-driven flux removal. They do, however, identify specific regions in which compression alone is insufficient to explain the observed evolution.

This proof-of-concept study demonstrates that density-based remote-sensing diagnostics can place CME-front evolution between an ideal expansion baseline and compression-driven snowplow benchmarks while identifying sectors that depart from both. The framework can be extended to larger event samples spanning different coronal environments and upstream solar-wind conditions. Combining it with event-specific in situ measurements and global MHD modeling will provide independent constraints on magnetic connectivity, upstream structure, and the processes governing CME-front evolution during propagation.

\begin{acknowledgements}
This research was conducted as part of the IKY-Fulbright Partnership Award for Visiting Research Students. AV and EP were supported by NASA grant 80NSSC22K0970.
SS and SP acknowledge support by the ERC Synergy Grant ‘Whole Sun’ (GAN: 810218).
\end{acknowledgements}

\bibliographystyle{aa-note} 

\bibliography{bibliography}

\begin{appendix}

\section{Bootstrap Extrapolation and Sensitivity Analysis}

\begin{figure*}[]
  \centering
  \includegraphics[width=2\columnwidth]{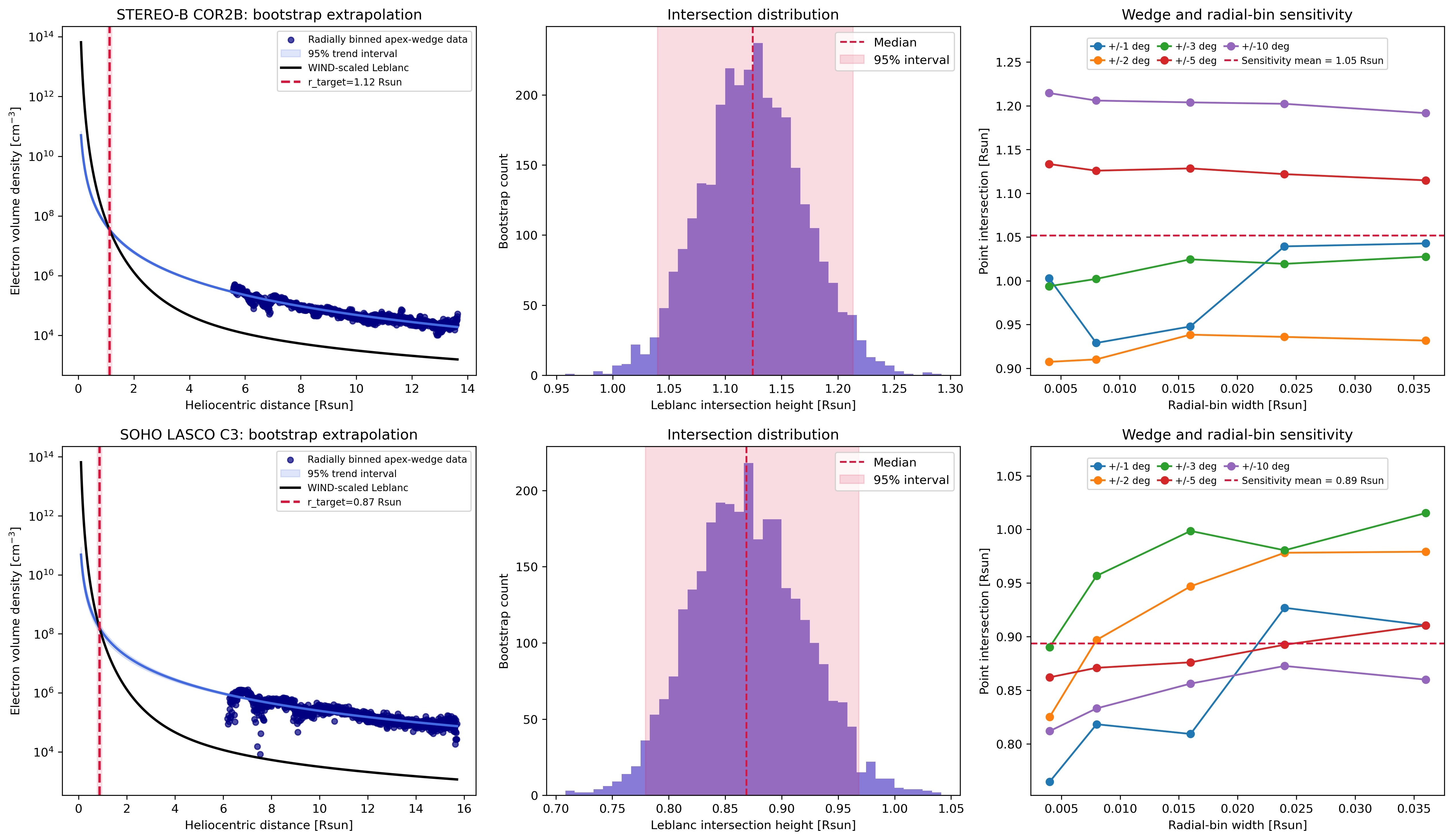}
  \caption{Bootstrap extrapolation and sensitivity analysis for STEREO-B/COR2B (top) and SOHO/LASCO C3 (bottom). The dashed red line in the sensitivity panels marks the mean intersection height across the tested wedge and radial-bin combinations.}
  \label{fig:opt_data2}
\end{figure*}

The same bootstrap extrapolation and sensitivity analysis applied to COR2-A is repeated here for COR2-B and LASCO C3. Figure~\ref{fig:opt_data2} shows, for each instrument, the radially binned apex-wedge density profile and its extrapolated trend, the bootstrap distribution of the intersection height with the WIND-scaled Leblanc background, and the sensitivity of this height to the adopted PA wedge and radial-bin widths. In both data sets, the analysis provides a consistency check on the adopted reference height and shows how the inferred intersection varies with the sampling parameters.

\clearpage

\section{Remote-Sensing Observations from COR2-A, COR2-B, and LASCO C3}

\begin{figure*}[]
  \centering
  \includegraphics[width=2\columnwidth]{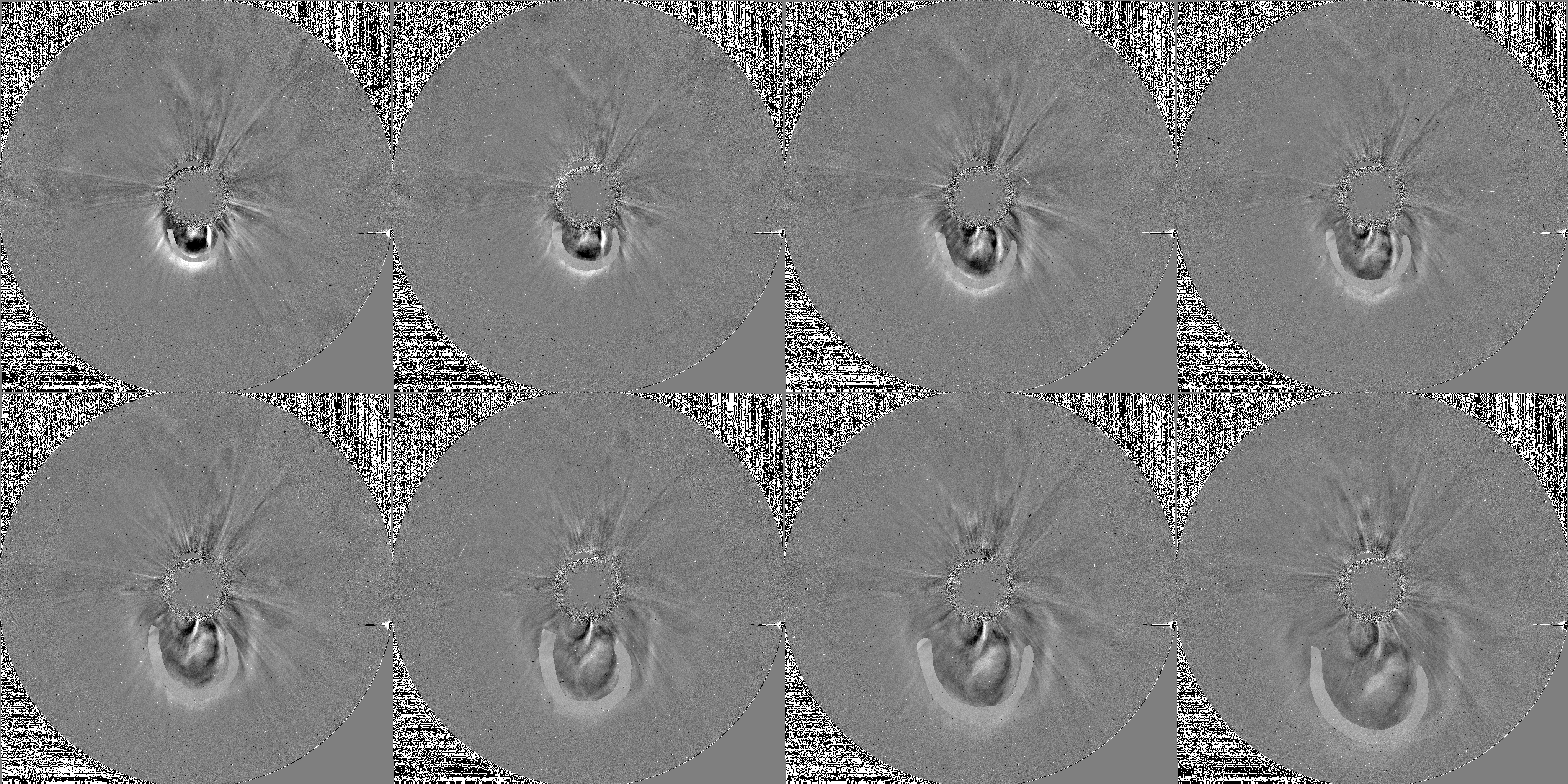}
  \caption{COR2-A remote-sensing observations of the 27 January 2012 CME with the analysis region of interest (ROI) overplotted. }
  \label{fig:obs_cor2a}
\end{figure*}


\begin{figure*}[]
  \centering
  \includegraphics[width=2\columnwidth]{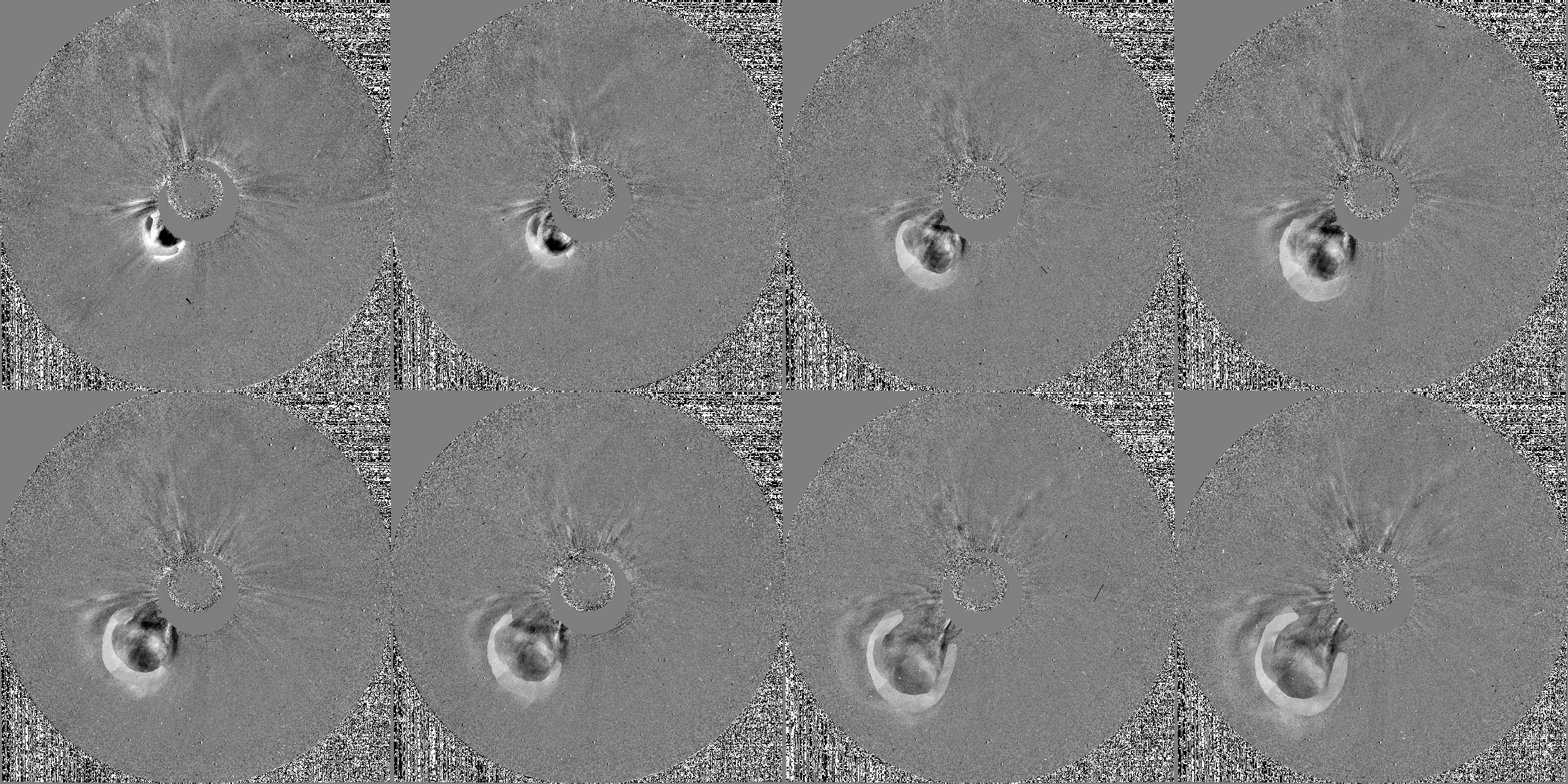}
  \caption{COR2-B remote-sensing observations of the 27 January 2012 CME with the analysis region of interest (ROI) overplotted. }
  \label{fig:obs_cor2b}
\end{figure*}


\begin{figure*}[]
  \centering
  \includegraphics[width=2\columnwidth]{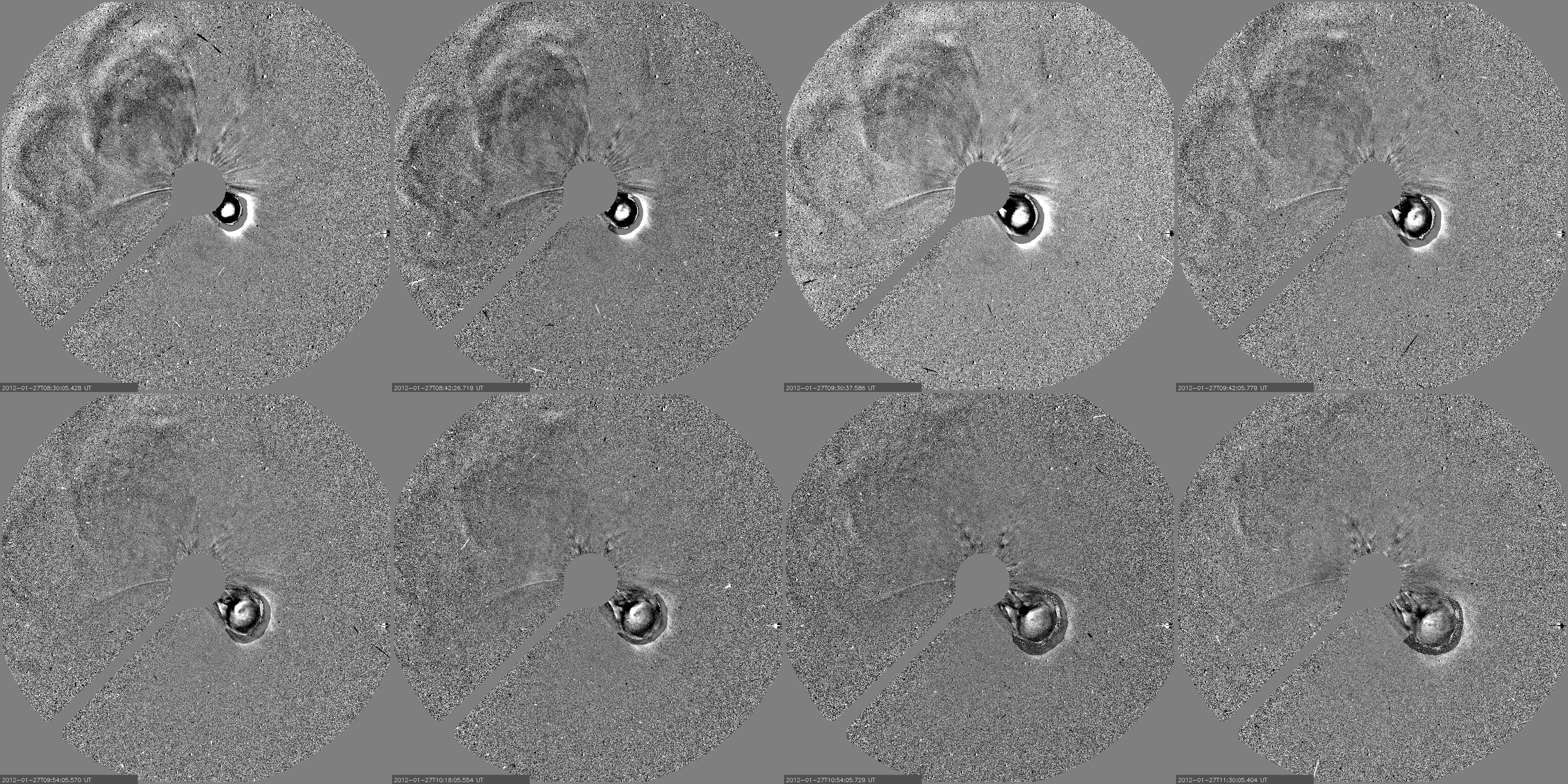}
  \caption{LASCO C3 remote-sensing observations of the 27 January 2012 CME with the analysis region of interest (ROI) overplotted. }
  \label{fig:obs_c3}
\end{figure*}

Figures~\ref{fig:obs_cor2a}--\ref{fig:obs_c3} show representative white-light coronagraph observations of the 27 January 2012 CME from COR2-A, COR2-B, and LASCO C3, with the analysis regions of interest (ROIs) overplotted. The ROIs outline the portion of the CME front used throughout this work to extract sector-resolved density profiles and to define the corresponding synthetic sampling for same-area comparisons. The inner and outer ROI boundaries bracket the bright front and provide a consistent definition of the front thickness and radial extent. The displayed ROIs are broader than the subset used for the quantitative analysis, because the front and its visibility evolve across the image sequence. For consistency, we base the analysis on the ROI subsection that is available in every timestamp, and we apply the same fixed subsection when sampling both the observations and the synthetic maps.

The three viewpoints provide complementary perspectives on the same evolving front. COR2-A offers the cleanest isolation of the front in this event and therefore serves as the primary reference in the main text. COR2-B and LASCO C3 provide a consistency check under different viewing perspectives and over larger heliocentric distances. In each case, the ROIs are placed to follow the coherent leading edge while minimizing contamination from background structures and trailing features, so that the extracted profiles primarily reflect the density evolution of the front relevant to pile-up and other transit-related modifications.

\clearpage

\section{Observations - Synthetic - ROI}

This appendix illustrates the spatial selection used for the observation--synthetic comparison. The manually defined observational ROI follows the leading-edge brightness enhancement, and the corresponding mask is applied to the synthetic image so that both density estimates sample the same projected portion of the CME front. The examples below provide a visual check of this spatial correspondence.

\begin{figure}[]
  \centering
  \includegraphics[width=\columnwidth]{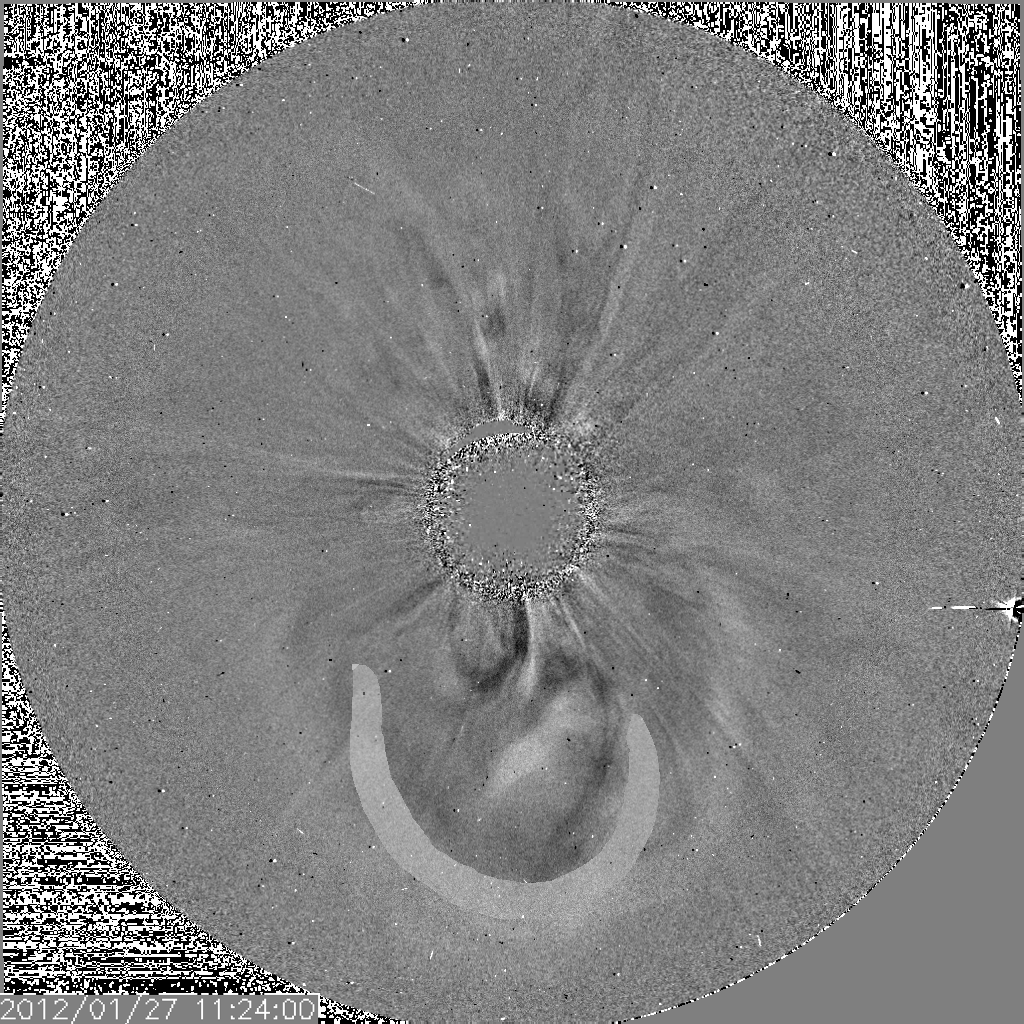}
  \caption{Representative white-light coronagraph observation with the manually defined ROI overplotted. The ROI brackets the leading-edge brightness enhancement used for the sector-resolved density extraction and is intended to capture the front region most relevant for pile-up, while potentially including part of the rear sheath.}
  \label{fig:obs}
\end{figure}

\begin{figure}[]
  \centering
  \includegraphics[width=\columnwidth]{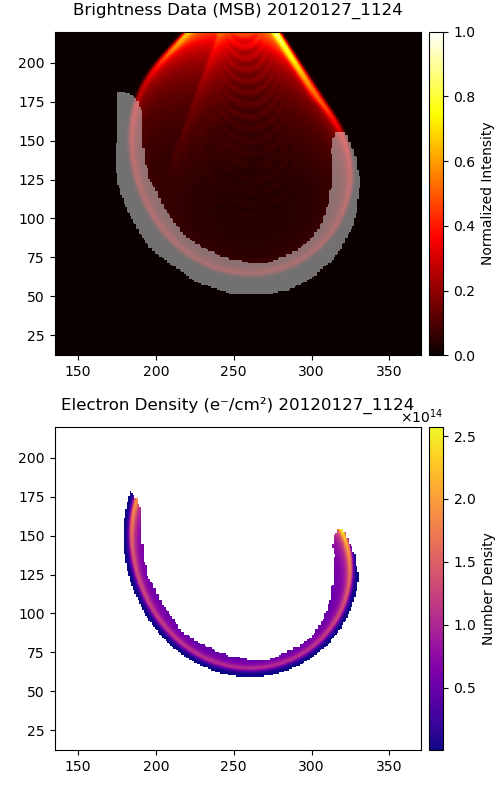}
  \caption{Synthetic GCS-based reference shell with the ROI overplotted (top) and the corresponding ROI-masked shell column density (bottom). The ROI mask is used to extract the synthetic front contribution within the same spatial selection adopted for the observations.}
  \label{fig:roi}
\end{figure}

\hfill\break
\clearpage

\section{COR2-B Results}
\label{Cor2b_results}

\begin{figure}[H]
  \centering
  \includegraphics[width=\columnwidth]{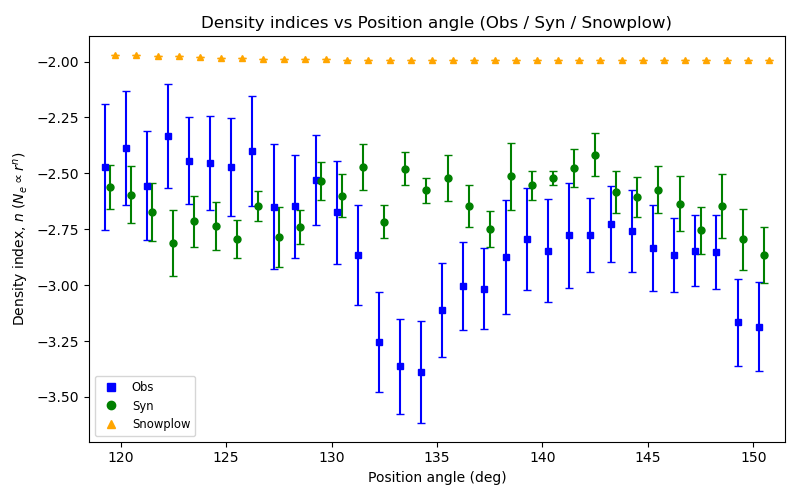}
  \caption{Radial density fall-off indices $n$ versus position angle for COR2-B, showing the observations (blue squares), the synthetic CME (green circles), and the snowplow model (orange triangles). Each point corresponds to a PA bin of $1^\circ$ and vertical bars show uncertainties on $n$. Flattening toward the snowplow value marks compression-dominated sectors, while indices more negative than the synthetic shell identify sectors with steeper-than-self-similar falloff.}
  \label{fig:density_index_vs_pa_cor2b}
\end{figure}

\begin{figure*}[!bp]
  \centering
  \includegraphics[width=\textwidth]{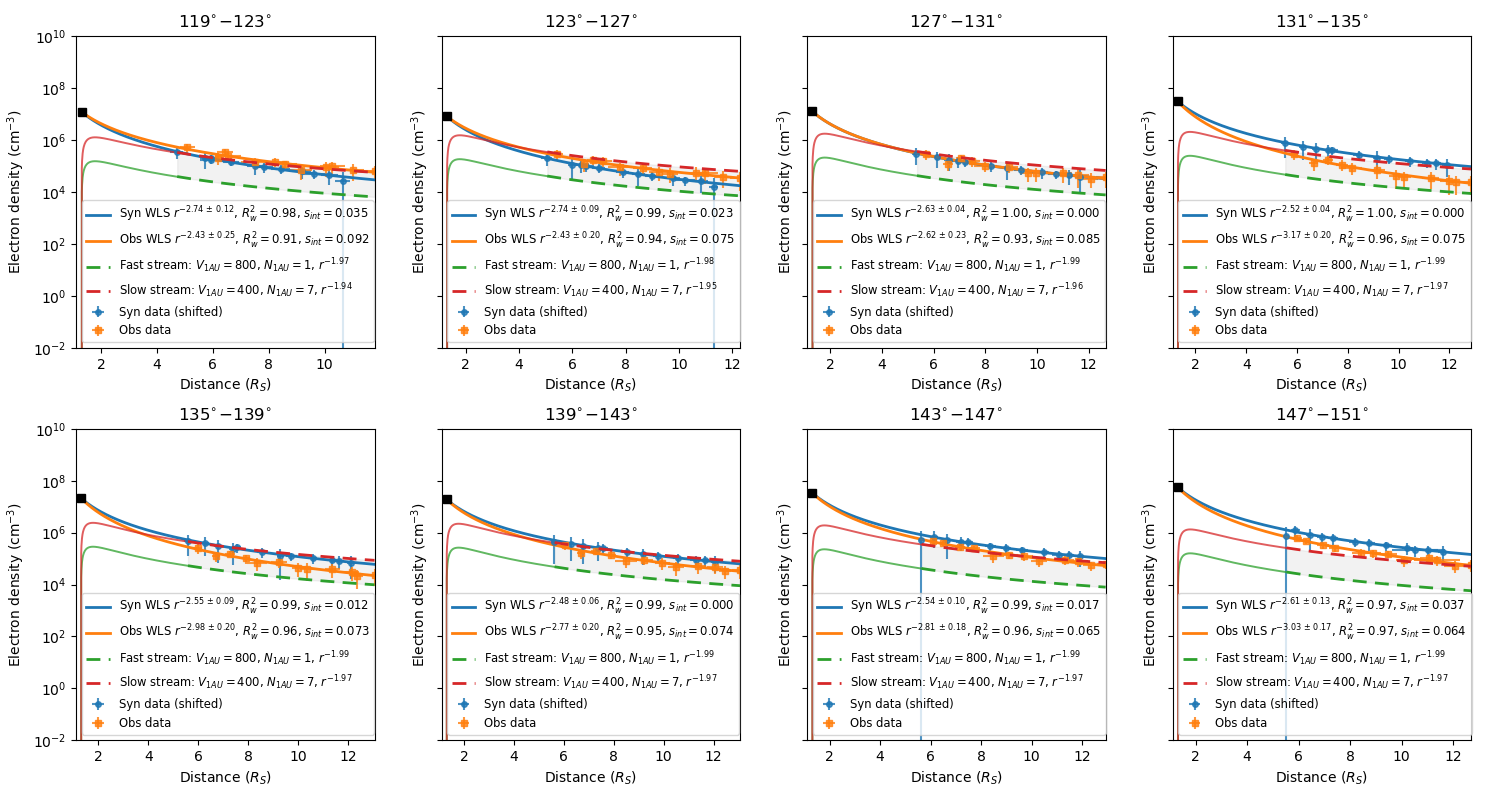}
  \caption{Sector-wise density comparison for COR2-B. Orange points show the observations, and blue points show the synthetic profiles normalized at $r_{\mathrm{target}}=1.3~R_{\mathrm{S}}$, marked by the black square. Error bars are based on the within-sector pixel standard deviations. Solid curves show weighted least-squares power-law fits, with the fitted exponents, $R_w^2$, and intrinsic scatter $s_{\mathrm{int}}$ reported in each panel. The fast- and slow-stream snowplow curves provide compression benchmarks for the adopted upstream conditions. Their representative power-law slopes are obtained by least squares over the radial span covered by the density profiles.}
  \label{fig:cor2b_sectors}
\end{figure*}

\hfill\break 
\hfill\break 

Figures~\ref{fig:cor2b_sectors} and
\ref{fig:density_index_vs_pa_cor2b} present the same sector-based
comparison shown for COR2-A in the main text, but for COR2-B.
Across most sectors, the observed COR2-B profiles decay more rapidly
with heliocentric distance than the self-similar synthetic shell,
indicating systematically steeper density declines relative to an
unperturbed expansion reference. The fitted synthetic slopes cluster
near the expected self similar behavior, with
$n_{\rm syn}\approx -2.52$ to $-2.74$, whereas the observed slopes
show greater variability, with $n_{\rm obs}\approx -2.43$ to $-3.17$
over the common fitting range.

The position-angle summary in
Figure~\ref{fig:density_index_vs_pa_cor2b} shows that the departure
from the synthetic reference is more pronounced in certain sectors,
particularly around $132^\circ$--$136^\circ$, but the pattern is less
systematic than in COR2-A. Several sectors show slopes close to the
synthetic reference, while others exhibit noticeably steeper declines.
Both the observed and synthetic slopes remain steeper than the
snowplow benchmark, which stays close to $n=-2$. This behavior is
consistent with spatial variations in the density evolution and with
line-of-sight effects that differ between viewpoints. Taken together,
the COR2-B results indicate that the density evolution across the
CME front is spatially structured, while also illustrating that the
detailed sector pattern inferred from density slopes can depend on
viewing perspective and on the ability to isolate the front cleanly
in a given instrument.

\clearpage

\section{LASCO C3 Results}
\label{LASCO-C3_results}

\begin{figure}[H]
  \centering
  \includegraphics[width=\columnwidth]{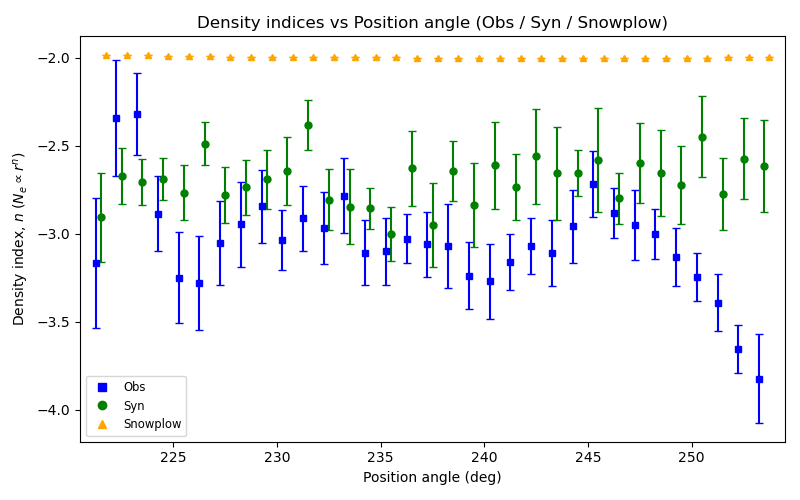}
  \caption{Radial density fall-off indices $n$ versus position angle for LASCO C3, showing the observations (blue squares), the synthetic CME (green circles), and the snowplow model (orange triangles). Each point corresponds to a PA bin of $1^\circ$ and vertical bars show uncertainties on $n$. Flattening toward the snowplow value marks compression-dominated sectors, while indices more negative than the synthetic shell identify sectors with steeper-than-self-similar falloff.}
  \label{fig:density_index_vs_pa_c3}
\end{figure}

\begin{figure*}[!bp]
  \centering
  \includegraphics[width=\textwidth]{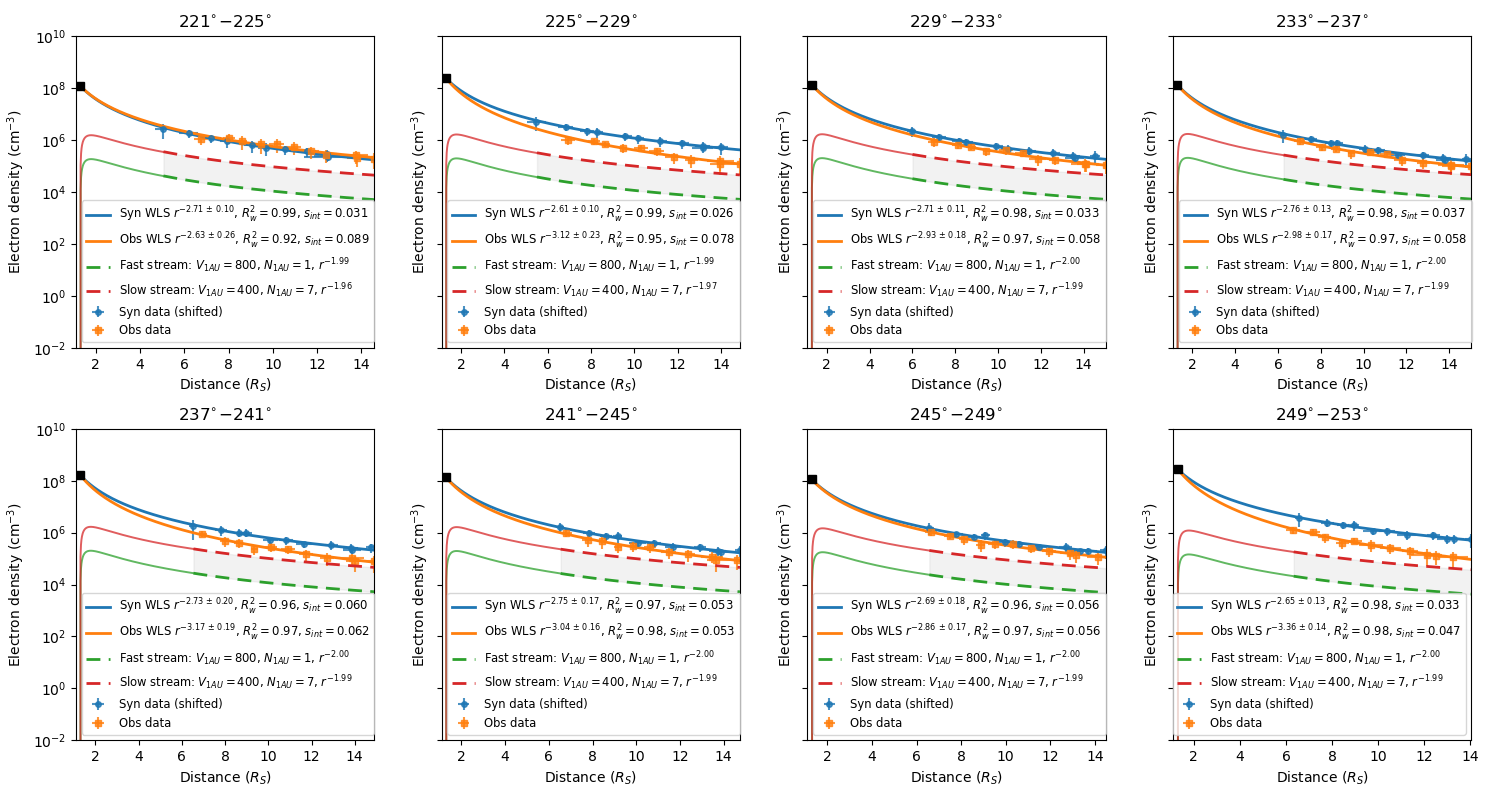}
  \caption{Sector-wise density comparison for LASCO C3. Orange points show the observations, and blue points show the synthetic profiles normalized at $r_{\mathrm{target}}=1.3~R_{\mathrm{S}}$, marked by the black square. Error bars are based on the within-sector pixel standard deviations. Solid curves show weighted least-squares power-law fits, with the fitted exponents, $R_w^2$, and intrinsic scatter $s_{\mathrm{int}}$ reported in each panel. The fast- and slow-stream snowplow curves provide compression benchmarks for the adopted upstream conditions. Their representative power-law slopes are obtained by least squares over the radial span covered by the density profiles.}
  \label{fig:c3_sectors}
\end{figure*}

\hfill\break 
\hfill\break 

We apply the same sector-resolved workflow used in the main text
to the LASCO C3 observations, extending the comparison to larger
heliocentric distances from the SOHO viewpoint.
Figure~\ref{fig:c3_sectors} shows that the observed C3 density
profiles generally decline more rapidly than the self-similar
synthetic shell. Over the common fitting ranges, the synthetic
slopes occupy a relatively narrow interval, with
$n_{\rm syn}\approx -2.76$ to $-2.61$, whereas the observed slopes
range from $n_{\rm obs}\approx -3.36$ to $-2.63$. The observed and
synthetic profiles closely match over $221^\circ$--$225^\circ$,
while the remaining displayed sectors have steeper observed
declines, with the largest difference over
$249^\circ$--$253^\circ$.

Figure~\ref{fig:density_index_vs_pa_c3} summarizes the azimuthal
variation at finer angular sampling. The observed indices are more
negative than the synthetic values across much of the sampled
front, although some position angles show agreement within the
uncertainties or locally flatter observed profiles. Both the observed
and synthetic indices remain steeper than the snowplow benchmark.
The C3 results therefore show that departures from the self-similar
reference remain spatially structured in the outer field of view.
Differences from the COR2 results cannot be attributed to radial
evolution alone, because the viewing geometry and projected sectors
also differ between instruments.

\end{appendix}

\end{document}